\pdfoutput=1
\documentclass[11pt,twoside,a4paper,cmspaper,final,collab]{cms-tdr}
\def\svnVersion{0f8ebe3}\def\svnDate{2022/09/19}\def\cmsCernNoTag{CERN-EP-2022-001}\def\cmsCernDate{\today}\def\cmsMessage{Published in the Journal of High Energy Physics as \href{http://dx.doi.org/10.1007/JHEP09(2022)088}{\doi{10.1007/JHEP09(2022)088}.}}
\begin{document}\cmsNoteHeader{B2G-20-002}

\DeclareRobustCommand{\PVLQT}{{\HepParticle{T}{}{}}\Xspace}
\newcommand{\TF}{\ensuremath{TF(\pt, \eta)}\xspace}
\newcommand{\TFV}{\ensuremath{TF_{\mathrm{v}}(\pt, \eta)}\xspace}
\newcommand{\tpr}{\ensuremath{\PVLQT}\xspace}
\newcommand{\bpr}{\ensuremath{\PB}\xspace}
\newcommand{\tZb}{\ensuremath{\PQt\PZ\PQb}\xspace}
\newcommand{\tHb}{\ensuremath{\PQt\PH\PQb}\xspace}
\newcommand{\mtzb}{\ensuremath{m_{\PQt\PZ\PQb}}\xspace}
\newcommand{\mthb}{\ensuremath{m_{\PQt\PH\PQb}}\xspace}
\newcommand{\msd}{\ensuremath{m_{\mathrm{SD}}}\xspace}
\newcommand{\msdt}{\ensuremath{m_{\mathrm{SD}}(\PQt)}\xspace}
\newcommand{\msdz}{\ensuremath{m_{\mathrm{SD}}(\PZ)}\xspace}
\newcommand{\msdh}{\ensuremath{m_{\mathrm{SD}}(\PH)}\xspace}
\newcommand{\itop}{\ensuremath{\text{imageTop}}\xspace}
\newcommand{\mdtop}{\ensuremath{\text{imageTop}_{\mathrm{MD}}}\xspace}
\newcommand{\db}{\text{Dbtag}\xspace}
\newcommand{\ztau}{\ensuremath{\tau_{21}}\xspace}
\newcommand{\cott}{\ensuremath{ \cot(\theta_2 )}\xspace}
\newcommand{\cottsq}{\ensuremath{ \cot^2(\theta_2 )}\xspace}
\newcommand{\st}{single \PQt}

\cmsNoteHeader{B2G-20-002}
\title{Search for a \texorpdfstring{\PWpr}{W\prime} boson decaying to a vector-like quark and a top or bottom quark in the all-jets final state at  \texorpdfstring{$\sqrt{s}=13\TeV$}{sqrt(s) = 13 TeV}}

\date{\today}

\abstract{
A search is presented for a heavy \PWpr boson resonance decaying to a \bpr or \tpr vector-like quark and a \PQt or a \PQb quark, respectively.  The analysis is performed using proton-proton collisions collected with the CMS detector at the LHC. The data correspond to an integrated luminosity of 138\fbinv at a center-of-mass energy of 13\TeV. Both decay channels result in a signature with a \PQt quark, a Higgs or \PZ boson, and a \PQb quark, each produced with a significant Lorentz boost. The all-hadronic decays of the Higgs or \PZ boson and of the \PQt quark are selected using jet substructure techniques to reduce standard model backgrounds, resulting in a distinct three-jet \PWpr boson decay signature.  No significant deviation in data with respect to the standard model background prediction is observed. Upper limits are set at 95\% confidence level on the product of the \PWpr boson cross section and the final state branching fraction.  A \PWpr boson with a mass below 3.1\TeV is excluded, given the benchmark model assumption of democratic branching fractions. In addition, limits are set based on generalizations of these assumptions.  These are the most sensitive limits to date for this final state.
}

\hypersetup{
pdfauthor={CMS Collaboration},
pdftitle={Search for a W' boson decaying to a vector-like quark and a top or bottom quark in the all-jets final state at sqrt(s) = 13 TeV},
pdfsubject={CMS},
pdfkeywords={CMS, vector-like quarks, substructure}}

\maketitle

\section{Introduction}
\label{sec:introduction}

New massive charged gauge bosons are predicted by various extensions of the
standard model (SM)~{\cite{doi:10.1146/annurev.nucl.55.090704.151502,PhysRevD.64.035002,PhysRevD.11.566}}.
The \PWpr boson is a hypothetical heavy partner of the SM \PW gauge boson that could be produced in
proton-proton ($\Pp\Pp$) collisions at the CERN LHC.  Searches for \PWpr bosons have
most recently been performed at a center-of-mass energy of 13\TeV by
the CMS and ATLAS Collaborations in the lepton-neutrino~\cite{wpCMS1,ATLAS:2019lsy}, diboson~\cite{Sirunyan:2018iff,wpAtlas1},
and diquark~\cite{wpCMS3,wpAtlas2} final states.
The decay of the \PWpr boson to a heavy \bpr or \tpr vector-like quark (VLQ) and a \PQt or \PQb quark,
respectively, is predicted, \eg, in composite Higgs boson models with custodial symmetry protection~\cite{AGASHE2005165,Barducci2013,Barducci2016}.
The VLQs are hypothetical heavy partners of SM quarks that contain left- and right-handed
chiralities, which transform identically under the SM gauge groups.
These models stabilize the quantum corrections to the Higgs boson mass and preserve naturalness.
The \PWpr branching fraction to a quark and a VLQ depends
on the VLQ mass, with a maximum of 50\%~\cite{Vignaroli:2014bpa}.  Searches for VLQs have been
performed by CMS and ATLAS in both the
single-~\cite{vlqs4,vlqs2,vlqs1,vlqs3,vlqs5} and pair-production~\cite{vlqp1,vlqp2,vlqp4,vlqp3} channels.

\begin{figure}[ht]
\centering
\includegraphics[width=0.45\textwidth]{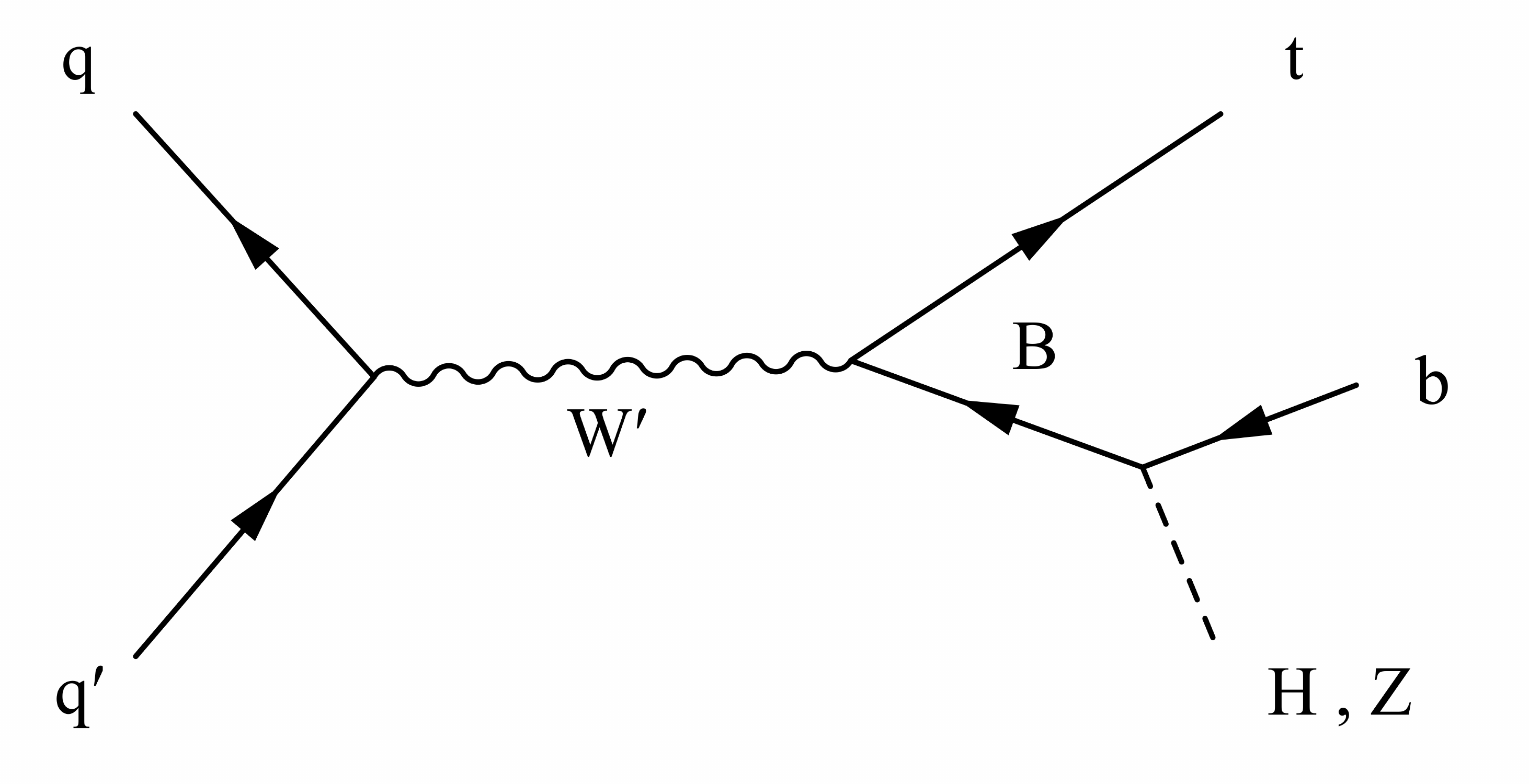}
\includegraphics[width=0.45\textwidth]{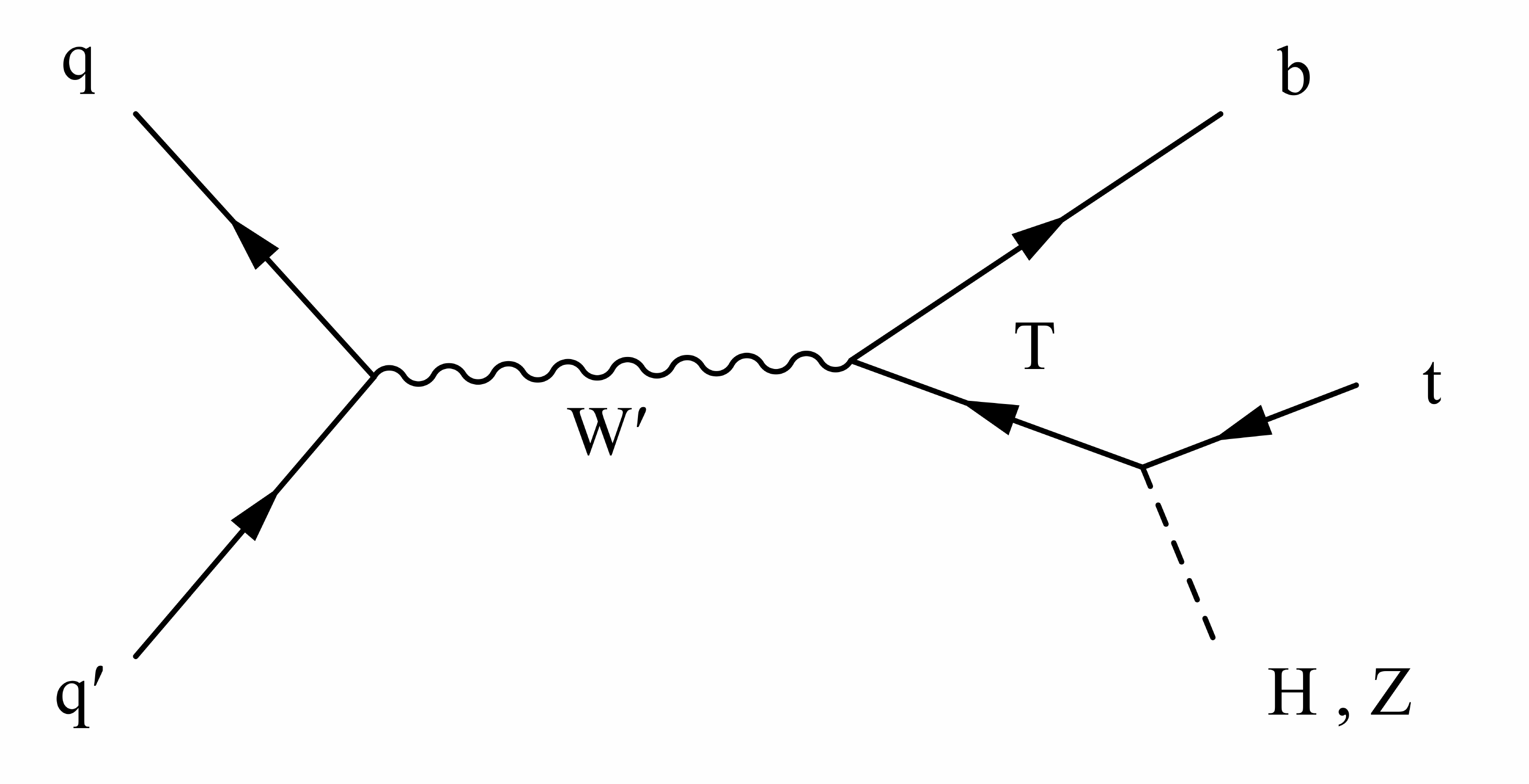}\\
\caption{
Dominant Feynman diagrams for the signal model considered.}
\label{figs:feyn}

\end{figure}

This paper describes a search for the \PWpr boson in the all-hadronic final state,
where the \bpr VLQ decays into a Higgs or \PZ boson and a \PQb quark, or the
\tpr VLQ decays into a Higgs or \PZ boson and a \PQt quark.
The separate studies of the two decay channels are referred to as the \tHb analysis
and the \tZb analysis throughout the paper.
Both the \bpr- and \tpr-mediated decays result in the same signature, as
shown in Fig.~\ref{figs:feyn}.

Owing to the high \PWpr boson and VLQ masses
considered, the decay products are highly Lorentz boosted, and are combined into
large-radius jets with distinct substructure.
The challenge is to distinguish between these events and the main background processes,
which arise from events comprised
of light-quark and gluon jets produced via the strong interaction, referred to as quantum chromodynamics (QCD)
multijet events, and \PQt quark pair (\ttbar) events.  These backgrounds are
modeled using a combination of Monte Carlo simulation and control regions in data.
The invariant mass distribution of the three-jet system (\mthb or \mtzb) is
used to search for the \PWpr boson signal.  The data correspond to
an integrated luminosity of 138\fbinv~\cite{CMS:2021xjt,CMS-PAS-LUM-17-004,CMS-PAS-LUM-18-002}
of $\Pp\Pp$ collisions at $\sqrt{s}=13\TeV$, recorded in the years 2016 to 2018.

We consider a theoretical framework where the \PQt quark and \PWpr
boson are superpositions of elementary and composite
modes, with the \PQt quark degree of compositeness given by $s_{\mathrm{L}}$, and the
mixing angle of the elementary and composite \PWpr states
given by $\theta_{2}$~\cite{Vignaroli:2014bpa}.  The \PWpr boson is assumed to be
produced in a Drell--Yan process, with a
cross section that is inversely proportional to \cottsq.  Small values of \cott tend to be dominated by the
leptonic \PWpr boson decay mode.  Large values of the $s_{\mathrm{L}}$ parameter
increase the relative phase space for the decay into two VLQs, whereas small
$s_{\mathrm{L}}$ values enhance the \PWpr diboson decays.  The analysis assumes
this theoretical framework as evaluated at $s_{\mathrm{L}}=0.5$ and $\cott=3$,
which is chosen to enhance sensitivity to the single VLQ decay channel of the \PWpr boson.
The expected signal cross sections are evaluated
at $\sqrt{s}=13\TeV$, considering \PWpr boson masses
in the 1.5 to 5.0\TeV range with the assumption that the $\PWpr\to\text{VLQ}$ branching
fraction is equally distributed between the $\PQt\bpr$ and $\PQt\tpr$ final states.
For the benchmark, the VLQ branching fractions for the decays
to $\PQq\PH$ and $\PQq\PZ$ are both assumed to be 50\%.  Additionally, we
present generalizations of the relative and total branching fraction assumptions for both
the $\PWpr$ and VLQ decays.

Tabulated results are provided in the HEPData record for this analysis~\cite{hepdata}.

\section{The CMS detector}
The central feature of the CMS apparatus is a superconducting solenoid of 6\unit{m} internal diameter,
providing a magnetic field of 3.8\unit{T}. Within the solenoid volume are a silicon pixel and strip tracker,
a lead tungstate crystal electromagnetic calorimeter (ECAL), and a brass and scintillator hadron calorimeter (HCAL),
each composed of a barrel and two endcap sections. Forward calorimeters extend the pseudorapidity ($\eta$) coverage provided by the
barrel and endcap detectors. Muons are detected in gas-ionization chambers embedded in the steel flux-return yoke outside the solenoid.

Events of interest are selected using a two-tiered trigger system. The first level,
composed of custom hardware processors, uses information from the calorimeters and muon
detectors to select events at a rate of around 100\unit{kHz} within a fixed latency of
about 4\mus~\cite{Sirunyan:2020zal}. The second level, known as the high-level trigger,
consists of a farm of processors running a version of the full event reconstruction
software optimized for fast processing, and reduces the event rate to around 1\unit{kHz}
before data storage~\cite{Khachatryan:2016bia}.

A more detailed description of the CMS detector, together with a definition of the coordinate system used and the
relevant kinematic variables, can be found in Ref.~\cite{Chatrchyan:2008zzk}.

The particle-flow algorithm~\cite{CMS-PRF-14-001} aims to reconstruct and identify each individual particle with an optimized
combination of information
from the various elements of the CMS detector. The energy of each photon is obtained from the ECAL measurement. The energy of each electron
is determined from a combination of the electron momentum at the
primary interaction vertex as determined by the tracker, the energy of the corresponding ECAL cluster, and the energy sum of all
bremsstrahlung photons spatially compatible with originating from the electron track. The energy of each muon is obtained from the momentum,
which is measured by the curvature of the corresponding track.  The energy of charged hadrons is determined from a combination of their momentum measured
in the tracker and the matching ECAL and HCAL energy deposits, corrected for zero-suppression effects and for the response function
of the calorimeters to hadronic showers. Finally, the energy of each neutral hadron is obtained from the corresponding corrected
ECAL and HCAL energies that are not associated with a charged hadron track.

Jets are clustered with the
anti-\kt algorithm~\cite{Cacciari:2008gp} using the \FASTJET 3.0~\cite{Cacciari:2011ma}
software package with distance parameters of 0.8 (AK8 jets) and 0.4 (AK4 jets).
Jet momentum is determined as the vectorial sum of all particle momenta in the
jet, and is found from simulation to be within
5 to 10\% of the true momentum over the entire \pt spectrum and detector acceptance.
Additional $\Pp\Pp$ interactions within
the same or nearby bunch crossings (pileup) can contribute additional tracks
and calorimetric energy depositions to the jet momentum.  To mitigate this
effect, tracks identified to be originating from pileup vertices are
discarded and an offset correction is applied to correct for remaining
contributions.

The candidate
vertex with the largest value of summed physics-object $\pt^2$ is taken to be
the primary $\Pp\Pp$ interaction vertex. The physics objects used for this determination are the jets,
clustered using the jet finding algorithm~\cite{Cacciari:2008gp,Cacciari:2011ma}
with the tracks assigned to candidate vertices as inputs, and the associated
missing transverse momentum, taken as the negative vector sum of the \pt of those jets.

To account for the neutral pileup component of the AK8 jets, the pileup
per particle identification (PUPPI)
algorithm~\cite{Bertolini2014} is used, which applies weights that rescale the
jet \pt based on the per-particle
probability of originating from the primary vertex prior to jet clustering.

Jet energy corrections are derived from simulation studies so that the average
measured response of jets becomes identical
to that of particle level jets. In situ measurements of the momentum balance in
dijet, $\text{photon} + \text{jet}$, $\PZ + \text{jet}$, and multijet events are
used to determine any residual differences between the jet energy scale (JES) in data
and in simulation, and appropriate corrections are made~\cite{Khachatryan:2016kdb}.
Additional selection criteria are applied to each jet to remove jets potentially
dominated by instrumental effects or reconstruction failures.
The jet energy resolution (JER) amounts typically to 15\% at 10\GeV, 8\% at 100\GeV,
and 4\% at 1\TeV, to be compared to
about 40, 12, and 5\% obtained when the calorimeters alone are used for jet clustering.

\section{Simulated samples}
\label{sec:datasample}
The \ttbar and \st quark production backgrounds are estimated from simulation generated at next-to-leading order with
\POWHEG 2.0~\cite{Nason:2004rx,Powheg,Alioli:2010xd,Alioli:2009je,Frederix:2012dh,Alioli:2011as}.  The signal
and QCD background events are generated at leading order using the
\MGvATNLO 2.3.3~\cite{Alwall:2014hca,Alwall:2007fs} package.  Signal events are generated in the
mass range from 1.5 to 5.0\TeV in 0.5\TeV increments.
Events are generated using either the NNPDF 3.0~\cite{Ball:2014uwa}
or 3.1~\cite{Ball:2017nwa} parton distribution functions (PDFs), with
parton showering and hadronization simulated with \PYTHIA 8.212~\cite{Sjostrand:2014zea}.

The simulated events are produced with either the CUETP8M2T4~\cite{CMS-PAS-TOP-16-021},
CUETP8M1~\cite{Khachatryan:2015pea}, or CP5~\cite{CMS:2019csb} underlying-event tunes.
The CMS detector is simulated using \GEANTfour~\cite{g4c}.
All simulated events include pileup simulation and are weighted such that the
distribution of the number of interactions per bunch crossing agrees with that in
data assuming a total inelastic cross section of 69.2\unit{mb}.

For each \PWpr boson mass ($m_{\PWpr}$) point, three characteristic VLQ mass points are generated,
corresponding to low (1/2 $m_{\PWpr}$), medium (2/3 $m_{\PWpr}$),
and high (3/4 $m_{\PWpr}$) mass ratios, in order to allow the full sensitive
phase space of the boosted \PWpr boson decay products to be explored.

The generated \PWpr boson and VLQ widths are chosen to be narrow
(less than 10\%) relative to the experiment resolution, as required by
the theoretical predictions for the region of phase space under investigation~\cite{Vignaroli:2014bpa}.

\section{Event reconstruction}
\label{sec:eventreconstruction}

The final state is characterized by three high-\pt jets from the
decay of the \PQt and \PQb quarks, and from either the Higgs or \PZ boson.
Because the signal of interest corresponds to a high mass resonance
decaying to multiple high-\pt
jets, data events are triggered by the presence of either a large value
of \HT (the scalar sum of all AK4 jet \pt in the event), an AK8 jet with a large \pt,
or an AK8 jet with a large jet mass.

The efficiency of the trigger selection is studied using a sample of events that
are triggered by at least one high-\pt muon.
The fraction of these events that pass the full trigger selection is defined as
the trigger efficiency.  The offline event selection
requires that the \HT is larger
than 1\TeV and the maximum jet mass is larger than 140\GeV, which ensures
that the trigger is close to fully efficient.
The small inefficiency (less than 1\%) due to the trigger selection is taken
into account as an event weight when processing simulated events.

The jets from the \PQt quark (\PQt jet) and from the Higgs or \PZ boson (Higgs or \PZ jet)
decay tend to be wide, massive, and have a distinct substructure, whereas the
jet from the \PQb quark (\PQb jet) tends to be narrower and have a lower
mass.  Therefore, an AK8 jet with $\pt>400\GeV$ is required for the Higgs or \PZ boson
candidate, and similarly for the \PQt quark
candidate, together with an AK4 jet with
$\pt>200\GeV$ for the \PQb candidate.
The angular separation $\Delta R = \sqrt{\smash[b]{(\Delta\phi)^2+(\Delta\eta)^2}}$, where $\phi$ is the
azimuthal angle, between the two AK8
jets is required to be at least 1.6 to
reduce the confusion of jet shapes arising from the abutment of jet boundaries,
which can bias the background estimate.  The AK8 jets are then selected as being
consistent with a \PQt quark or a Higgs or \PZ boson decay using the jet
identification (``tagging'') procedures
defined below.  The jet considered for the \PQb quark candidate
is the leading AK4 jet with a $\Delta R$ of at least 1.2 from the tagged AK8
jets.  All jets are required to be within the tracker acceptance ($\abs{\eta}<2.4$).
The high-\pt requirement for the AK8 jets provides a strong reduction in the
background for the selection of boosted resonances. However, for the lowest mass
\PWpr boson signal hypothesis considered, it results in a small kinematic acceptance,
thus this mass region would also benefit from nonboosted techniques that are
beyond the scope of this paper.

\label{sec:Trigger}

\subsection{Tagging \texorpdfstring{\PQt}{t} jets}
\label{sec:toptag}

For \PQt quarks with $\pt>400\GeV$, the decay products, one \PQb quark and two light quarks, can
merge into a single AK8 jet. The \PQt quark jets are identified by a novel
algorithm designed to separate \PQt jets from jets arising from light quarks and gluons using a
deep convolutional neural network called \itop~\cite{Sirunyan:2020lcu}.
The network uses the particle momentum and particle-flow identification
category as an analogy to the
pixel intensity and color that would be used as inputs to a conventional
image-classification algorithm.  Additionally, the algorithm includes \PQb tagging
information simultaneously in the training in order to identify the \PQb quark
from the \PQt quark decay.

The version used in this analysis includes a
network that is trained on QCD jets that are drawn from a mass distribution
that is constrained to be identical to the top mass distribution, so as to
decorrelate the discriminator from the jet mass, and is referred to as the mass-decorrelated \itop or \mdtop.  This analysis is the
first application of the \mdtop tagger, which is the most sensitive
decorrelated \PQt tagger developed by the CMS Collaboration,
and the mass decorrelation is valid over a large range of \PQt jet mass
and \pt.
This tagger
shows an improvement of approximately a factor of six in background rejection
at a constant efficiency when compared to the best \PQt quark taggers used previously.
The \mdtop tagger is used to
select \PQt quark jets at an operating point with $\approx$60\% efficiency and a mistag probability of $\approx$2\% in QCD multijet events.

Owing to the decorrelated nature of the \PQt tagging algorithm,
we also apply a selection on the mass of the \PQt jet.
The modified mass-drop tagger algorithm~\cite{Dasgupta2013},
also known as the ``soft-drop'' (SD) algorithm~\cite{Larkoski:2014wba}, with $\beta=0$
and $z=0.1$, is used to calculate the mass variable \msdt to discriminate between
\PQt jets and the typically lighter background jets.  This variable is corrected by applying AK4 jet energy corrections to the
subjets identified by the algorithm. For events in the
signal region, we require $140<\msdt<220\GeV$, to be consistent with the
reconstructed mass of a \PQt quark.

A simulation-to-data scale factor for the \PQt tagging requirement is extracted
by fitting the SD mass distribution in a highly enriched sample of SM \ttbar events in data and simulation.
This scale factor is applied
to correct differences between the top tagging efficiencies for simulated events and data.  The scale factor is \pt dependent with a typical value of $\approx$0.95 and
does not vary significantly between different running periods.

\subsection{Tagging Higgs jets}
\label{sec:htag}
In the case of a highly boosted Higgs boson in the $\bbbar$ decay mode, the
decay products tend to merge into a single jet with a mass that is
consistent with that of a Higgs boson and contains two \PQb hadron decays.
The SD algorithm is used to extract the
variable \msdh as a measure of the jet mass, but in this case, the
jet is scaled using a $\pt$- and $\eta$-dependent correction suitable for
two-prong resonances below the \PQt quark mass~\cite{PhysRevD.97.072006}.

To be selected, the jet should fulfill the condition $105<\msdh<140\GeV$, to be consistent with the reconstructed Higgs boson mass.
Scale factors are used for the jet mass scale and resolution (JMS and JMR), which are derived
from a fit to the SD mass distribution of the \PW boson in
a sample enriched in lepton+jets \ttbar events using the technique
outlined in Ref.~\cite{CMS-PAS-JME-16-003}.

To identify the two \PQb hadrons clustered into the merged Higgs jet, a dedicated
double-\PQb tagging algorithm (\db) is used at an operating point
that has an efficiency of $\approx$60\% and a mistag probability of $\approx$7\% for \PQu, \PQd, \PQs quarks and gluon jets.
This algorithm aims to identify two \PQb quarks from the Higgs decay by using
variables correlated with \PQb hadron lifetime and mass combined using a boosted decision tree.
Data enriched in QCD-produced $\bbbar$ and \ttbar events are used to establish scale factors for \db for signal (which is $\approx$0.9)
and mistagged \PQt quarks, respectively~\cite{Sirunyan:2017ezt}.

\begin{figure}[htb]
\centering
\includegraphics[width=0.45\textwidth]{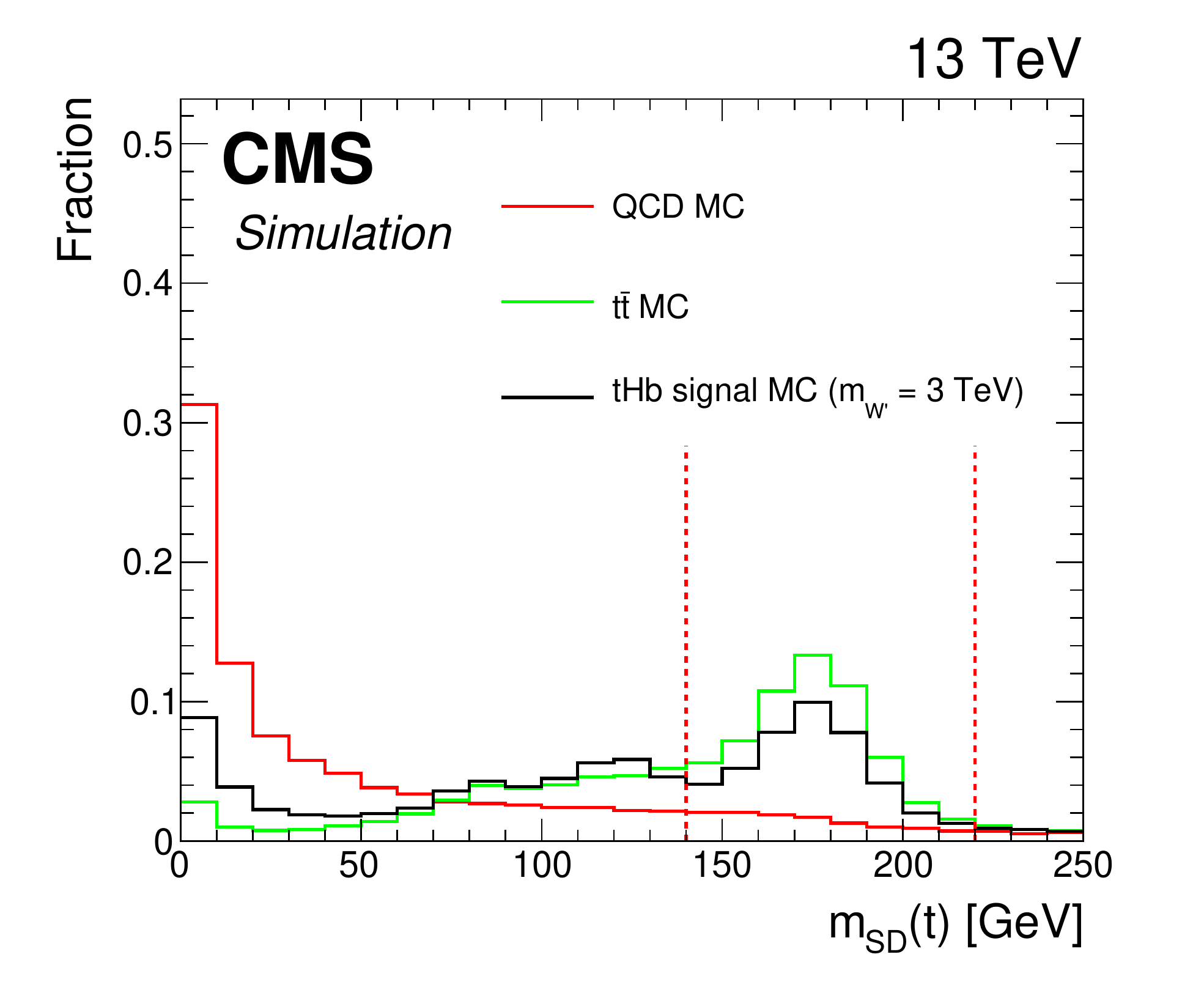}
\includegraphics[width=0.45\textwidth]{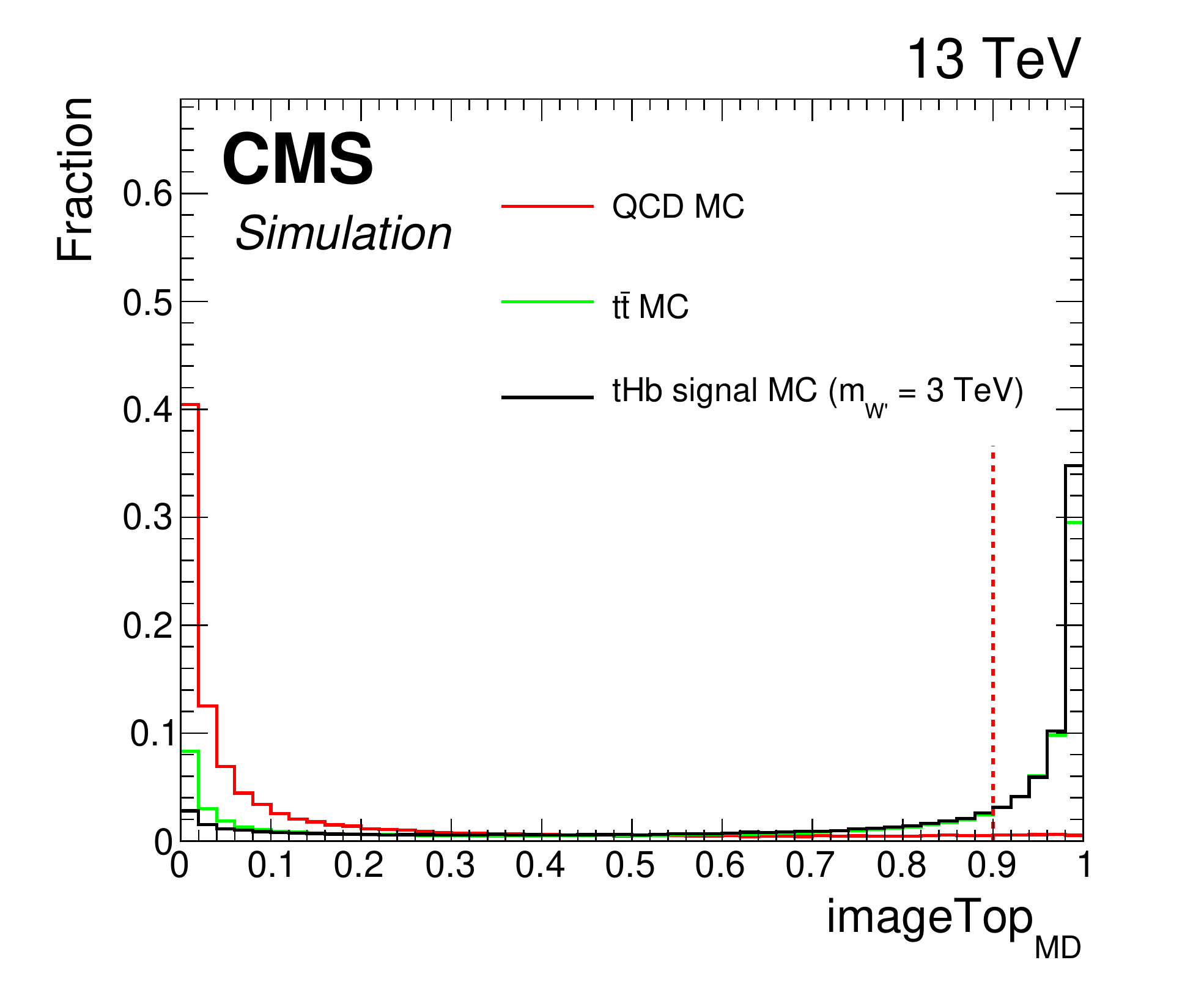}\\
\includegraphics[width=0.45\textwidth]{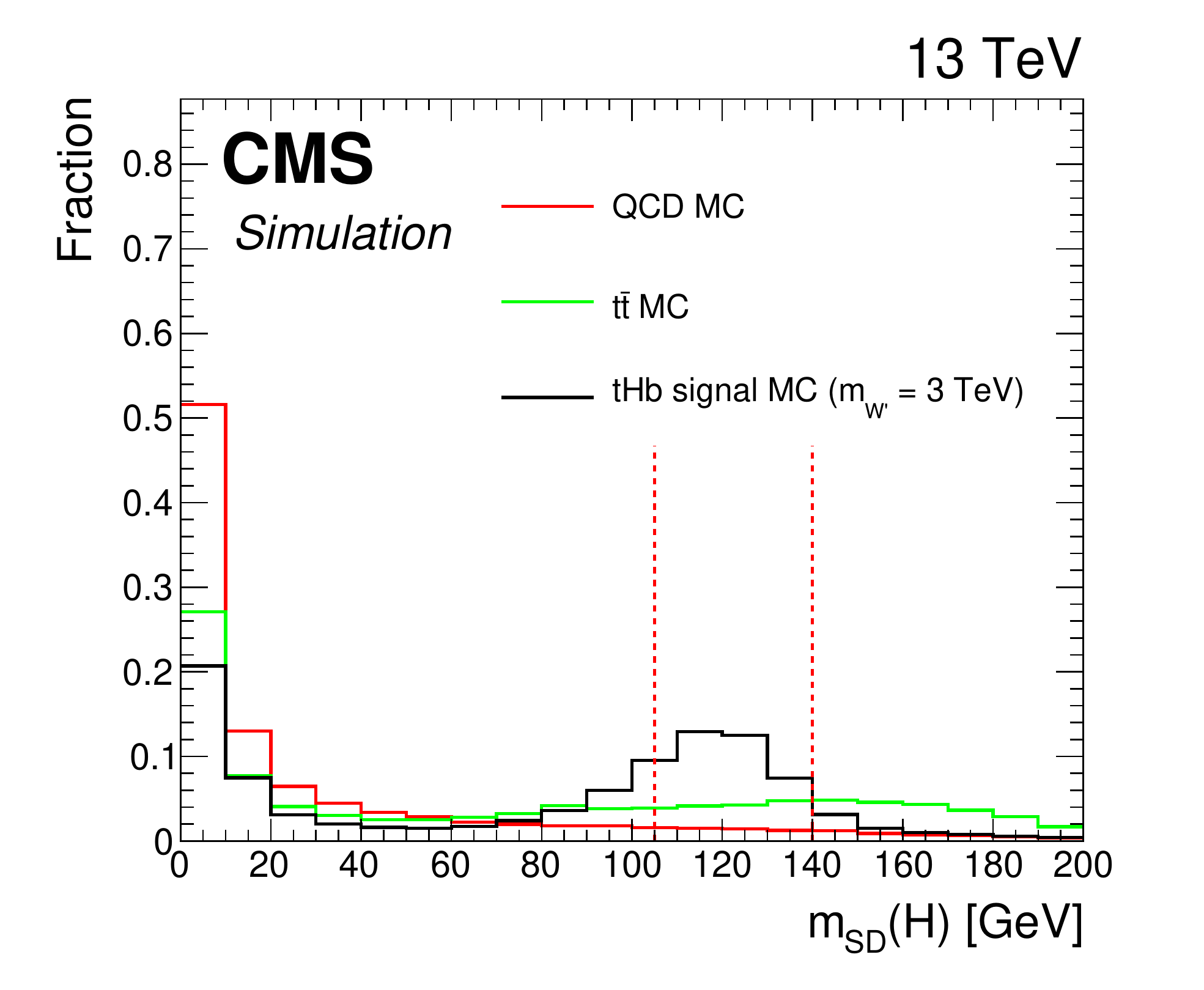}
\includegraphics[width=0.45\textwidth]{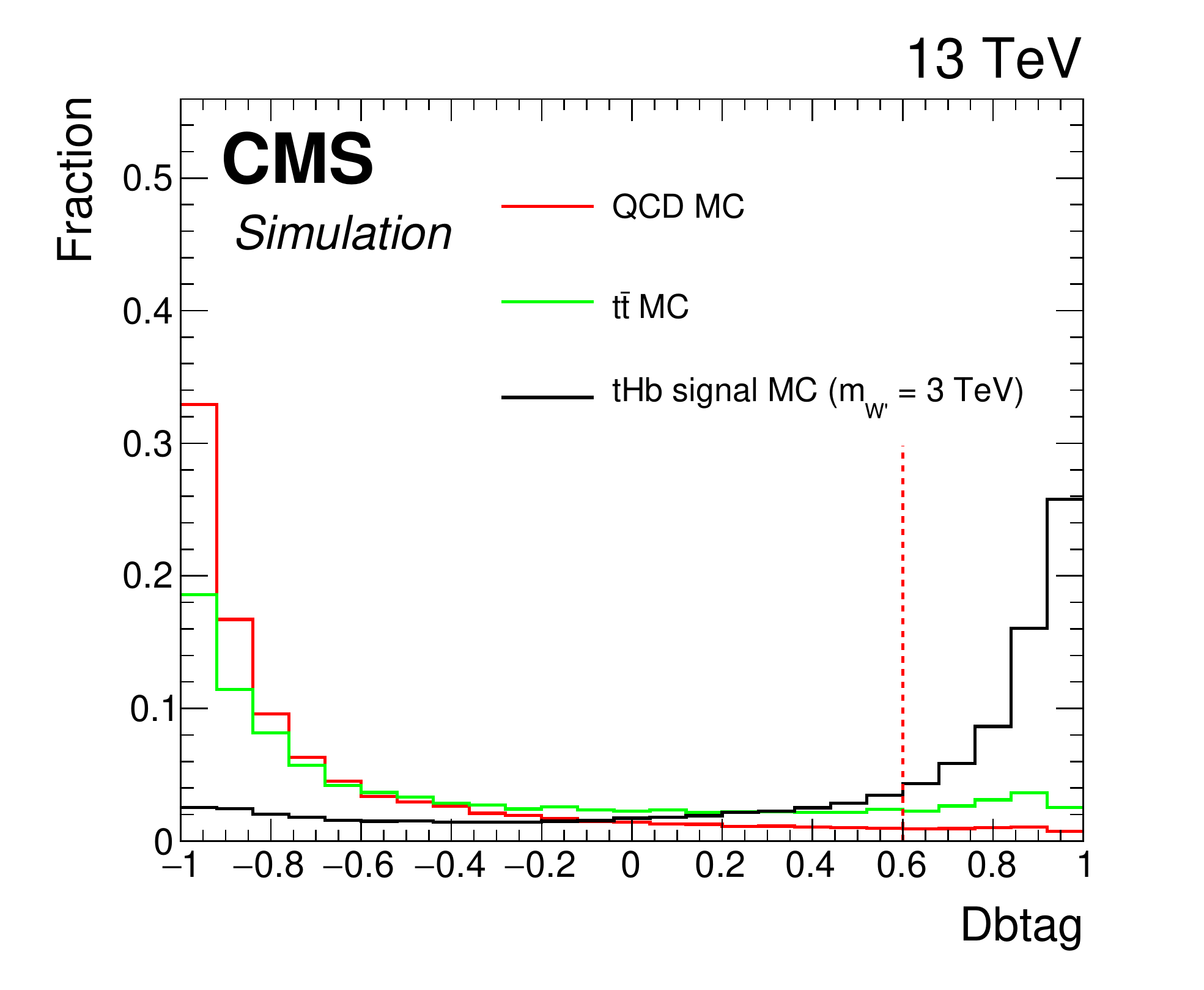}\\
\caption{
Simulated distributions of the discriminating variables for \ttbar, QCD, and
\tHb signal simulated events normalized to unity for the \tHb analysis.  Discriminant thresholds are shown as vertical dashed lines.
}
\label{figs:NM1plotsthb}
\end{figure}

Figure~\ref{figs:NM1plotsthb} shows the
simulated distributions of discriminants that are used for \PQt and Higgs jet
tagging in \ttbar, QCD, and signal for the \tHb analysis.

\subsection{Tagging \texorpdfstring{\PZ}{Z} jets}
\label{sec:ztag}

In contrast to the Higgs boson, the \PZ boson has significant decay
branching fractions to all five kinematically accessible quark flavors, so we require
a jet energy deposition pattern that is consistent with coming from two hard partons,
along with a jet mass that is consistent with a \PZ boson.

The $N$-subjettiness~\cite{Thaler:2011gf} algorithm defines the $\tau_{N}$ variable, which quantifies the consistency of the jet energy pattern with the expectation from
$N$ or fewer hard partons.  The $N$-subjettiness is defined as
\begin{linenomath}
\begin{equation}
	\tau_{N} = \frac{1}{d}\sum_{i}{{\pt}_i \min\{\Delta R_{1,i},\Delta R_{2,i},\ldots,\Delta R_{N,i} \}},
\end{equation}
\end{linenomath}
where the sum is over jet constituents.  The $\tau_{N}$ discriminant is bound between zero and one, with low values being more consistent with $N$ or fewer partons.  For a
\PZ boson hadronic decay, the ratio of $\tau_2$ to $\tau_1$ (\ztau) is used
at an operating point of $\approx$85\% efficiency and a mistag probability of $\approx$20\% in QCD multijet events.
Scale factors used for the $N$-subjettiness selection are derived
simultaneously with those for the JMS and JMR as discussed in Section~\ref{sec:htag}, and
have a value of $\approx$0.98.

The \msdz is required to be within the range $65<\msdz<105\GeV$, to be
consistent with the mass of a \PZ boson.

\begin{figure}[htbp]
\centering
\includegraphics[width=0.45\textwidth]{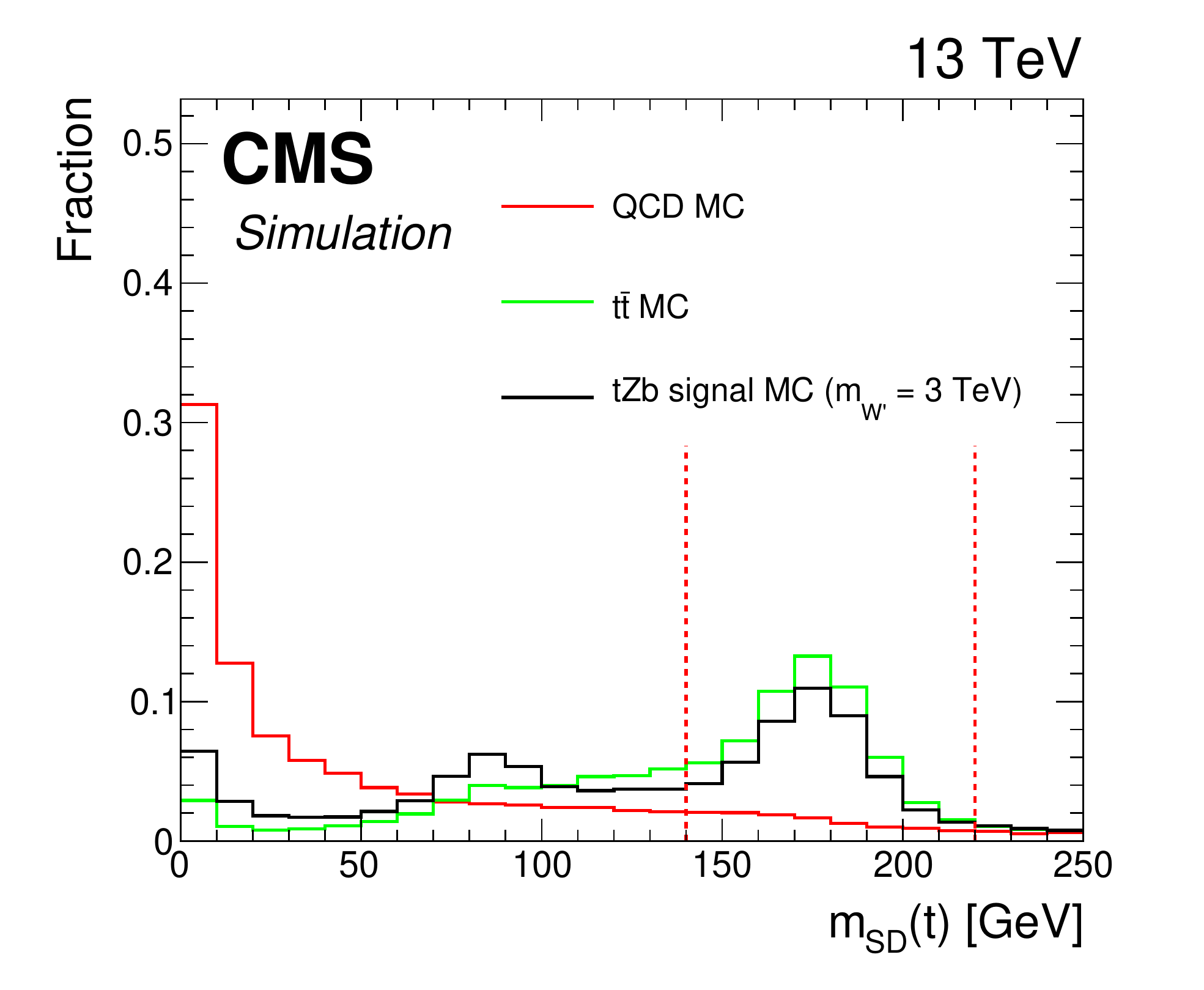}
\includegraphics[width=0.45\textwidth]{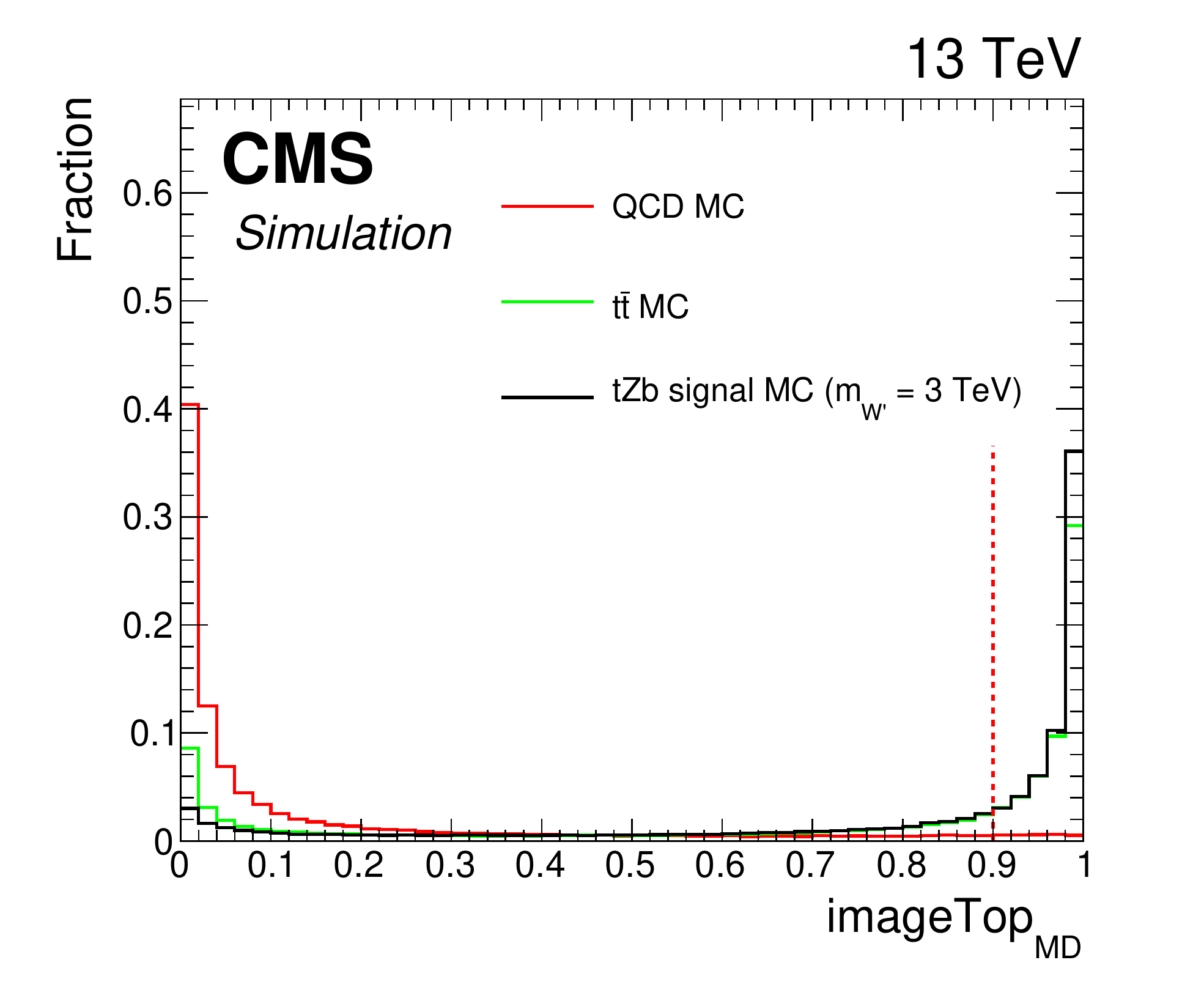}\\
\includegraphics[width=0.45\textwidth]{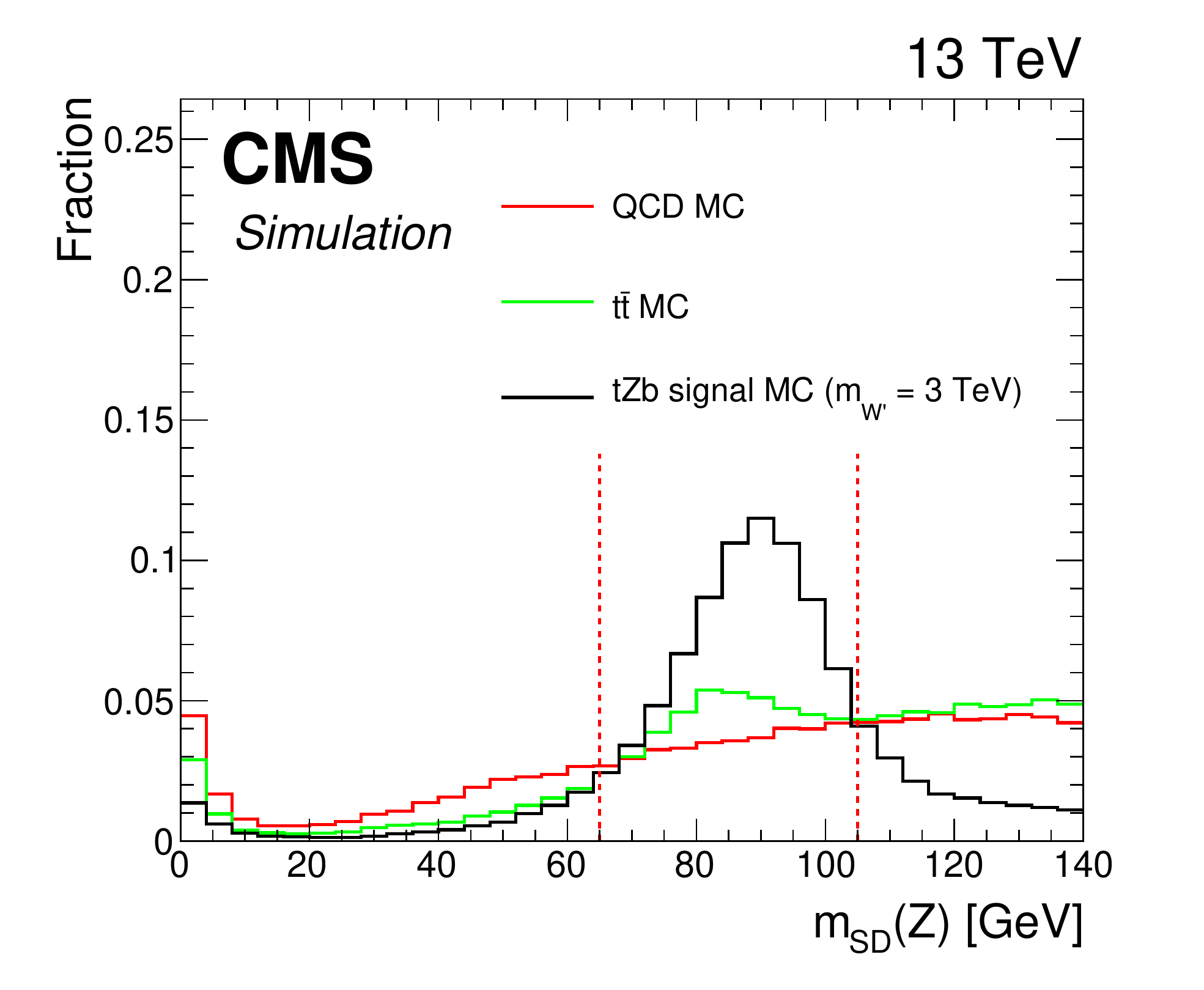}
\includegraphics[width=0.45\textwidth]{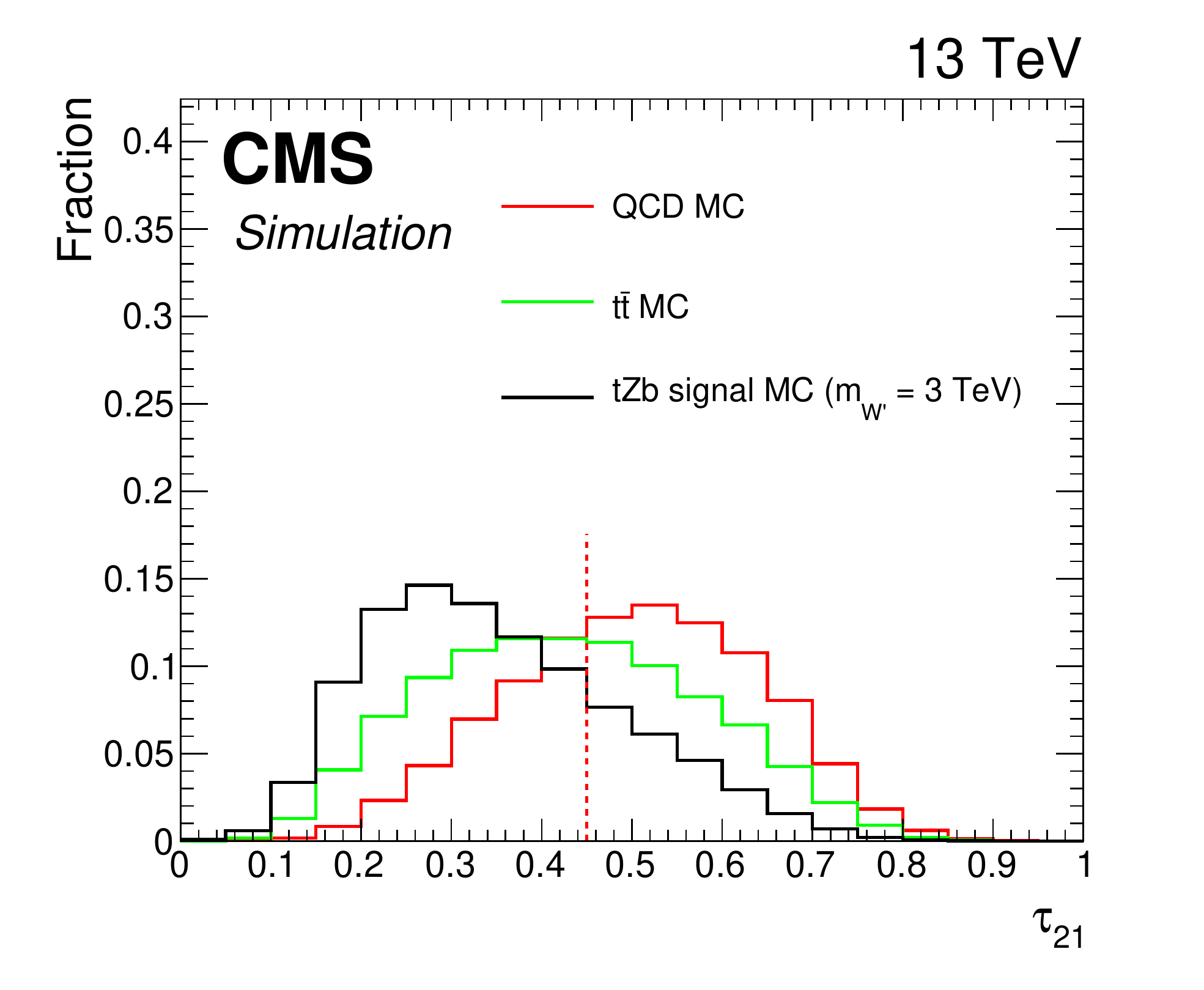}\\
\caption{
Simulated distributions of the discriminating variables for \ttbar, QCD, and
\tZb signal simulated events normalized to unity for the \tZb analysis.  Discriminant thresholds are shown as vertical dashed lines.
}
\label{figs:NM1plotstzb}
\end{figure}

Figure~\ref{figs:NM1plotstzb} shows the
distributions of discriminants that are used for \PQt and \PZ jet
tagging in \ttbar, QCD, and signal simulation for the \tZb analysis.

\subsection{Tagging \texorpdfstring{\PQb}{b} jets}
The \PQb quark from the VLQ or \PWpr boson decay is reconstructed as an AK4 jet
that is required to pass the \textsc{DeepJet} \PQb tagging algorithm~\cite{Bols:2020bkb,CMS-DP-2018-058}
at an operating point defined by a 1\% light-quark and gluon misidentification probability,
approximately 10\% \PQc quark misidentification probability, and approximately 80\% \PQb quark efficiency.
A simulation-to-data scale factor for the \PQb tagging requirement is used to
improve the agreement between data and simulation and has a value of $\approx$0.95.

\subsection{Event selection}
We categorize events into selection regions that are either used
to search for the \PWpr boson signal, predict the background in the signal region, or
validate this background estimate.
Event selection details are presented in Table~\ref{table:regions}.  The signal
region is required to contain a tight \PQt jet, a Higgs or \PZ jet, and a \PQb jet.
Regions used for background estimation and validation require combinations of
loosened tagger criteria.

The sensitivity of the selections has been studied both in
the context of the expected cross section upper limit and of the \PWpr boson discovery potential.  After
identifying the \PQt, Higgs or \PZ, and \PQb jets, the \PWpr boson candidate
mass is taken as the invariant mass of the three jets.  Table~\ref{table:sigeff}
shows the signal efficiency for all samples considered.

\begin{table}[htbp]
\topcaption{Selection regions used for signal identification and background estimation.  The AK8 jet discriminant and mass
selections are explicitly defined here for the \PQt, Higgs, and \PZ jet tags.
}
\centering
\label{table:regions}
\begin{tabular}{lccc}
Label & Tag & Discriminant & Mass\\
\hline
Tight & \PH & $0.6<\db$  &  $105<\msdh<140\GeV$\\
 & \PZ & $\ztau<0.45$  &  $65<\msdz<105\GeV$\\
 & \PQt & $0.9<\mdtop$  &  $140<\msdt<220\GeV$\\ \\
Medium & \PH & $0<\db<0.6$  &  $105<\msdh<140\GeV$\\
 & \PZ & $0.45<\ztau<0.6$  &  $65<\msdz<105\GeV$\\
 & \PQt & $0.3<\mdtop<0.9$  &  $140<\msdt<220\GeV$\\ \\
Inverted & \PH & $-1<\db<0$  &  $5<\msdh<30\GeV$\\
 & \PZ & $0.6<\ztau<1$  &  $5<\msdz<30\GeV$\\
 & \PQt & $0<\mdtop<0.3$  &  $30<\msdt<65\GeV$\\
\end{tabular}
\end{table}

\begin{table}[htb]
\topcaption{The signal efficiency (in percent) for the three VLQ mass ranges
considered.  The efficiency is given for the \tHb and \tZb final states,
separated into the low, medium, and high VLQ mass categories.}
\label{table:sigeff}
\centering
\begin{tabular}{ccccccc}
$m_{\PWpr}(\GeVns)$ & \tHb low & \tHb medium & \tHb high &  \tZb low & \tZb medium  &  \tZb high\\
\hline
1500 & 0.29 & 0.34 & 0.28 & 0.43 & 0.51 & 0.39 \\
2000 & 1.2 & 1.4 & 0.78 & 1.7 & 1.9 & 1.1 \\
2500 & 2.0 & 1.9 & 1.3 & 2.8 & 2.7 & 1.8 \\
3000 & 2.3 & 2.2 & 1.8 & 2.7 & 3.0 & 2.5 \\
3500 & 2.3 & 2.2 & 1.8 & 3.2 & 3.0 & 2.5 \\
4000 & 2.2 & 2.0 & 1.8 & 3.1 & 2.9 & 2.6 \\
4500 & 2.0 & 1.9 & 1.7 & 2.8 & 2.7 & 2.5 \\
5000 & 1.8 & 1.7 & 1.5 & 2.6 & 2.5 & 2.4 \\
\end{tabular}
\end{table}

\section{Background estimation}
\label{sec:backgroundestimation}

\begin{figure}[htb]
\centering
\includegraphics[width=0.9\textwidth]{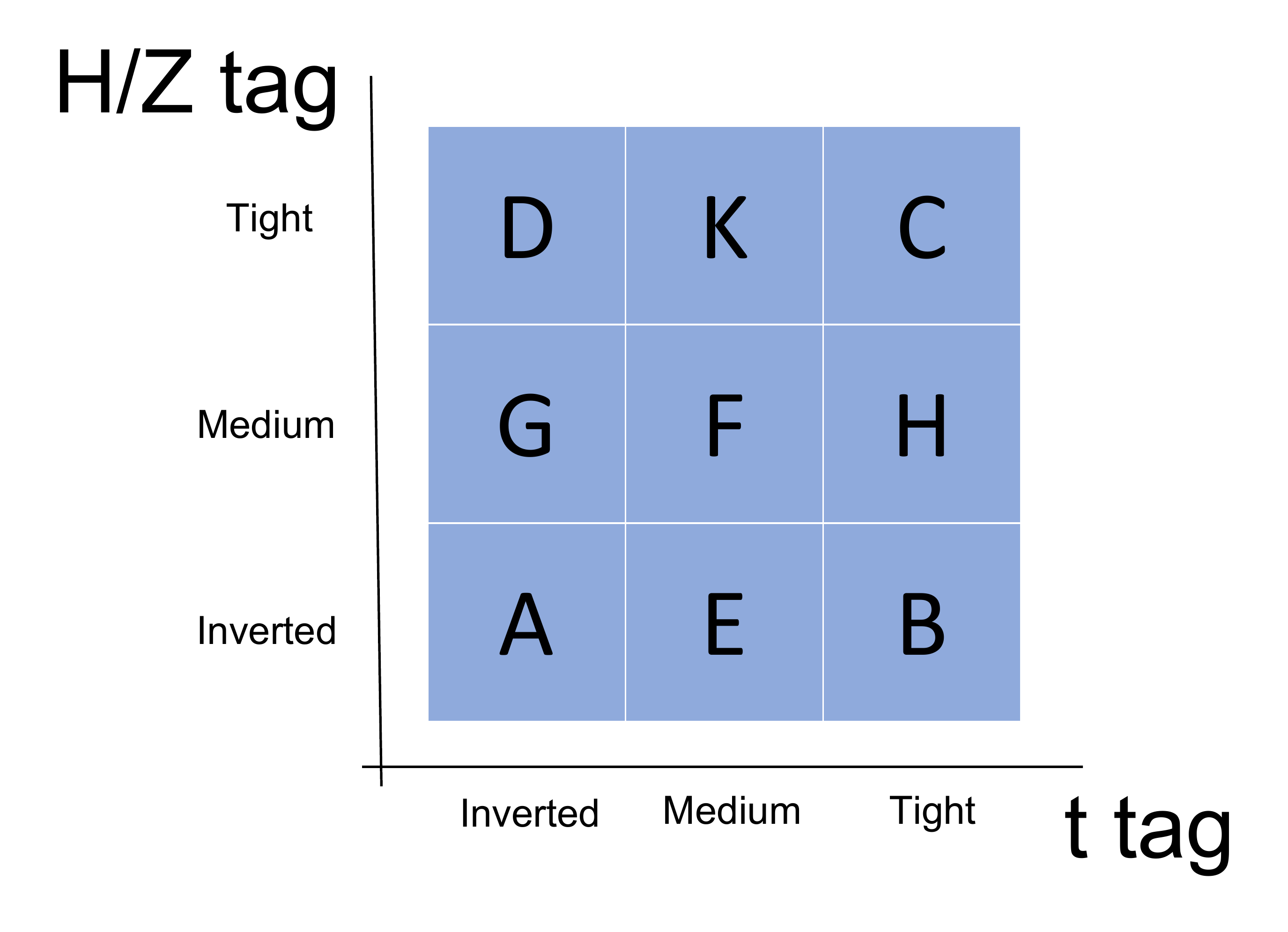}\\
\caption{
Diagram showing the regions used for background estimation.
The signal region is C, the regions A and B are used to determine \TF, and F, K, and H are validation regions.
The $x$ axis indicates the \PQt tag category, and the $y$ axis represents the
Higgs or \PZ boson tag category. The inverted, medium, and tight tag category definitions are given in Table~\ref{table:regions}.}
\label{figs:cartoon}
\end{figure}

The primary background is from QCD multijet production and is estimated from data.
For this estimate, we use control regions that are
selected with identical kinematic criteria to the signal
region, but with a reduced acceptance for signal events.  Table~\ref{table:regions} and
Fig.~\ref{figs:cartoon} define various selection regions.  A transfer function \TF is extracted from data by inverting the
Higgs or \PZ jet candidate selection.  After this inversion, \TF is defined as the ratio of the
jet \pt spectrum of the \PQt candidate in the case that it passes (region B)
to the case that it fails (region A) the \PQt tagging algorithm.
The \TF is extracted in two $\eta$ ranges (central,
$\abs{\eta}<1.2$, and forward, $\abs{\eta}>1.2$, regions).

The \TF distribution is used to predict the background in the signal region.
This is accomplished by defining the control region D, which has identical Higgs or \PZ and
\PQb jet candidate selections to the signal region, but with an inverted
\PQt jet selection.  Events that pass the region D selection use \TF as an event weight in a given
\PQt-candidate jet (\pt, $\eta$) bin.  These weighted events are then
used to provide the QCD background estimate in the signal region.  The primary assumption
for this background estimate method is that the \PQt jet substructure selection
can be inverted without significantly biasing the Higgs or \PZ jet substructure selection.

In the \TF extraction procedure, the \ttbar production component
derived from simulation is subtracted
from data to ensure that \TF
refers only to the QCD component.
Additionally, the \ttbar contamination of the QCD background
estimate in the signal region must be subtracted.  This is performed by
applying the QCD background estimation procedure to simulated \ttbar events
using the same \TF as is used when extracting the QCD estimate from data.
The following relations are used to extract the QCD component of the signal region:
\begin{linenomath}
\begin{equation}
\label{equation:regions}
\begin{split}
\TF &\equiv (\mathrm{B}_{\text{data}}-\mathrm{B}_{\ttbar})/(\mathrm{A}_{\text{data}}-\mathrm{A}_{\ttbar}), \\
\mathrm{C}_{\mathrm{QCD}} &\simeq (\mathrm{D}_{\text{data}}-\mathrm{D}_{\ttbar}) \TF.
\end{split}
\end{equation}
\end{linenomath}
Here, the subscripts ``data'' and ``\ttbar'' indicate that the distributions
are obtained from data and \ttbar events, respectively.

\begin{figure}[htbp]
\centering
\includegraphics[width=0.45\textwidth]{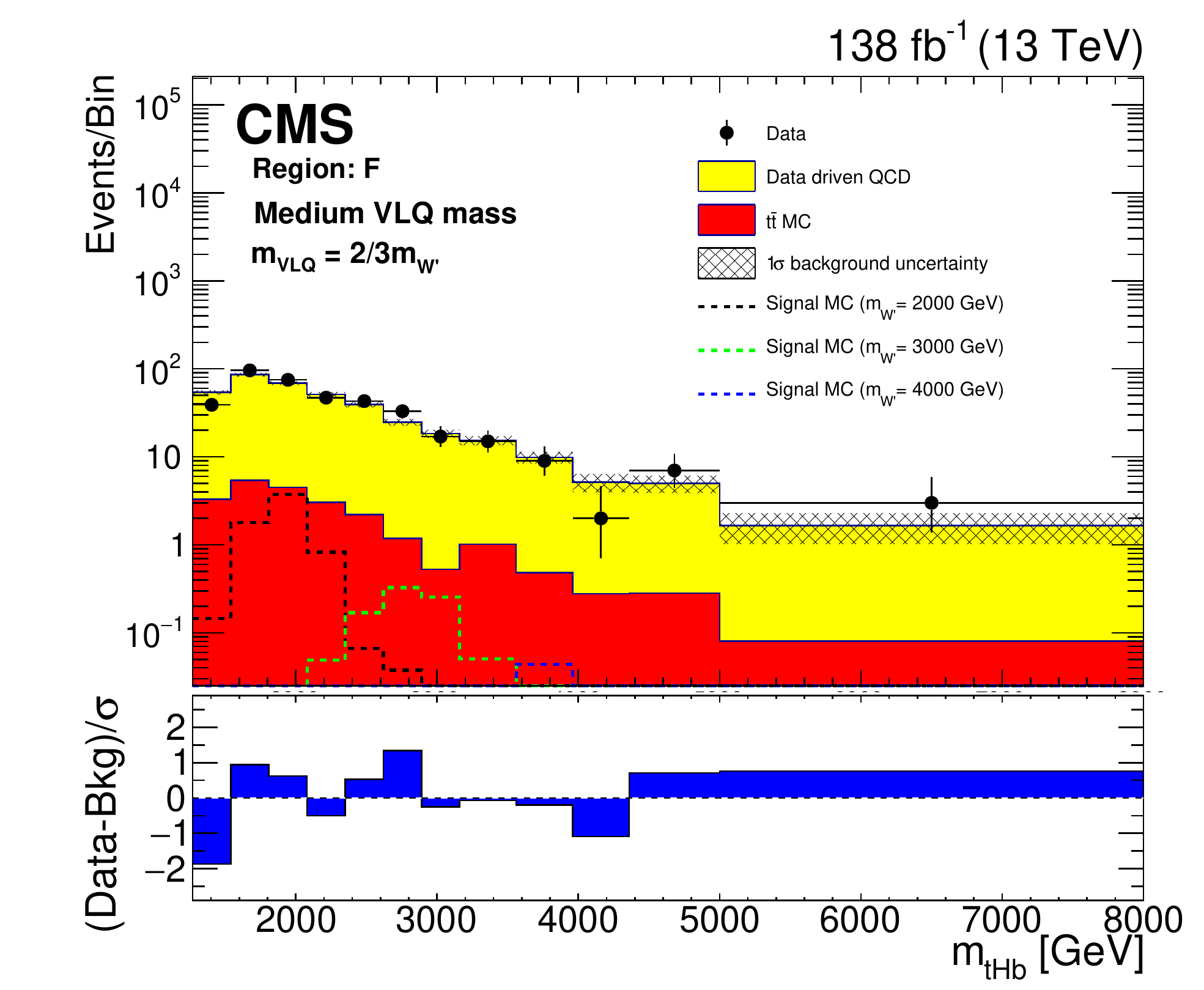}
\includegraphics[width=0.45\textwidth]{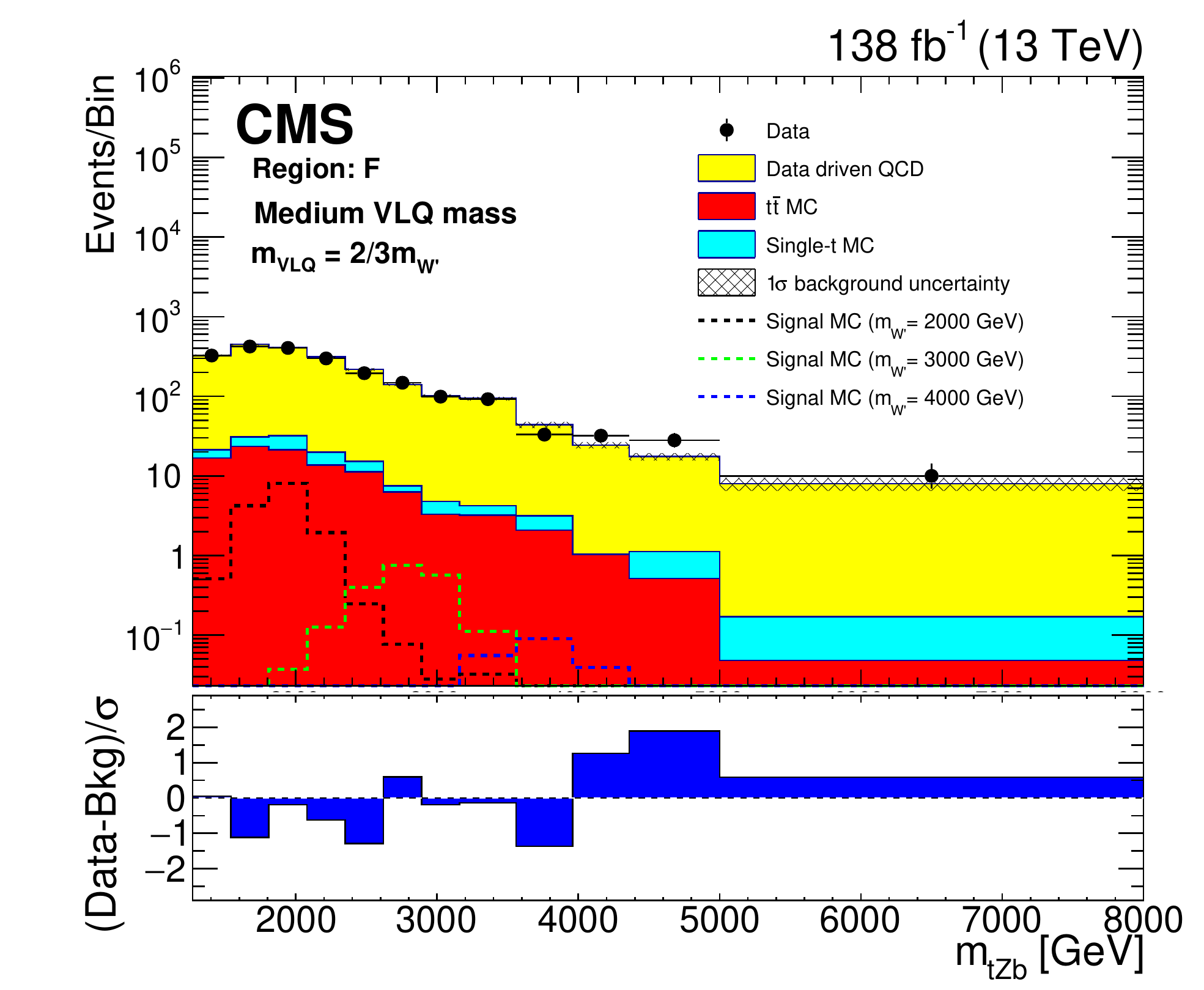}\\
\includegraphics[width=0.45\textwidth]{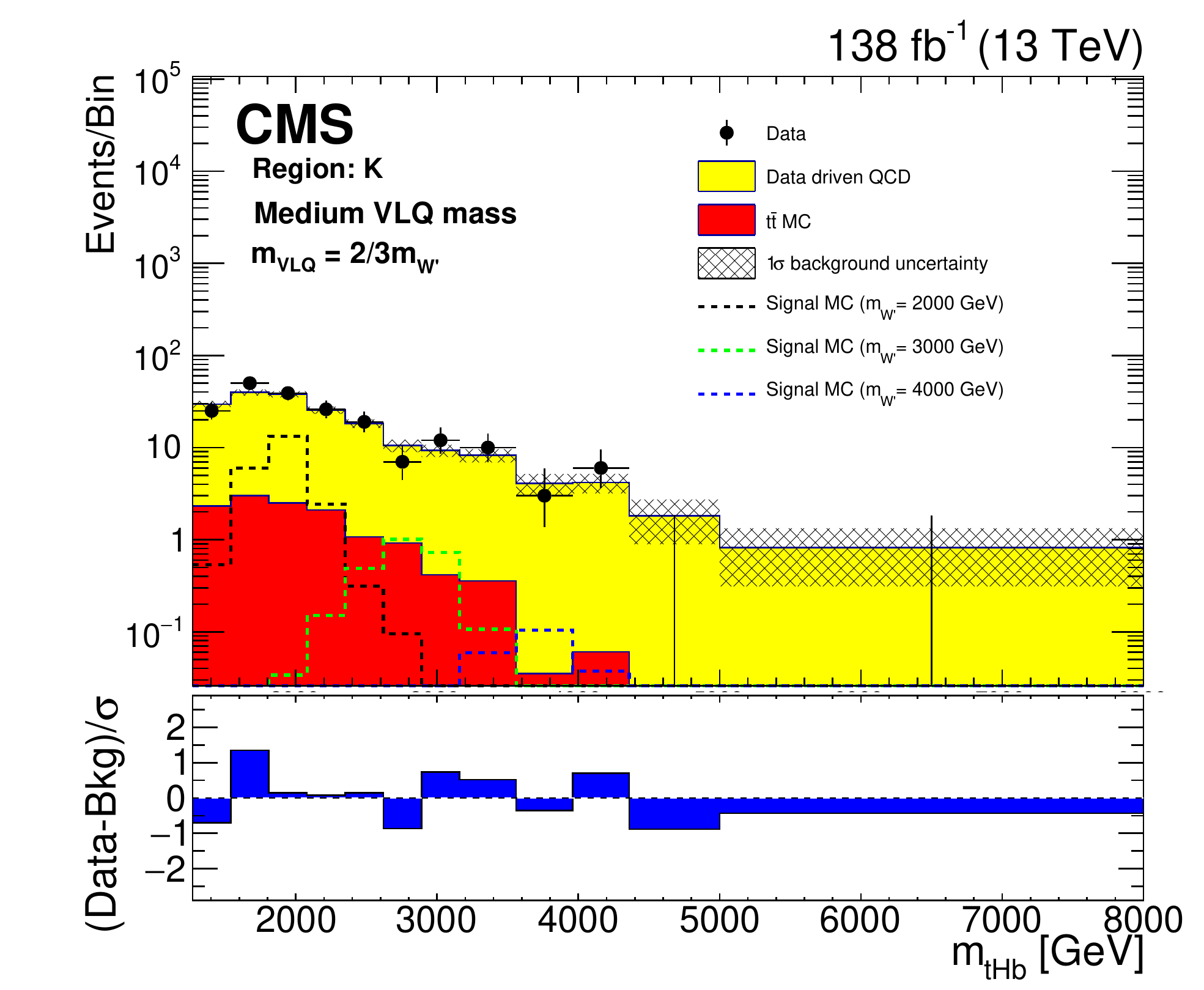}
\includegraphics[width=0.45\textwidth]{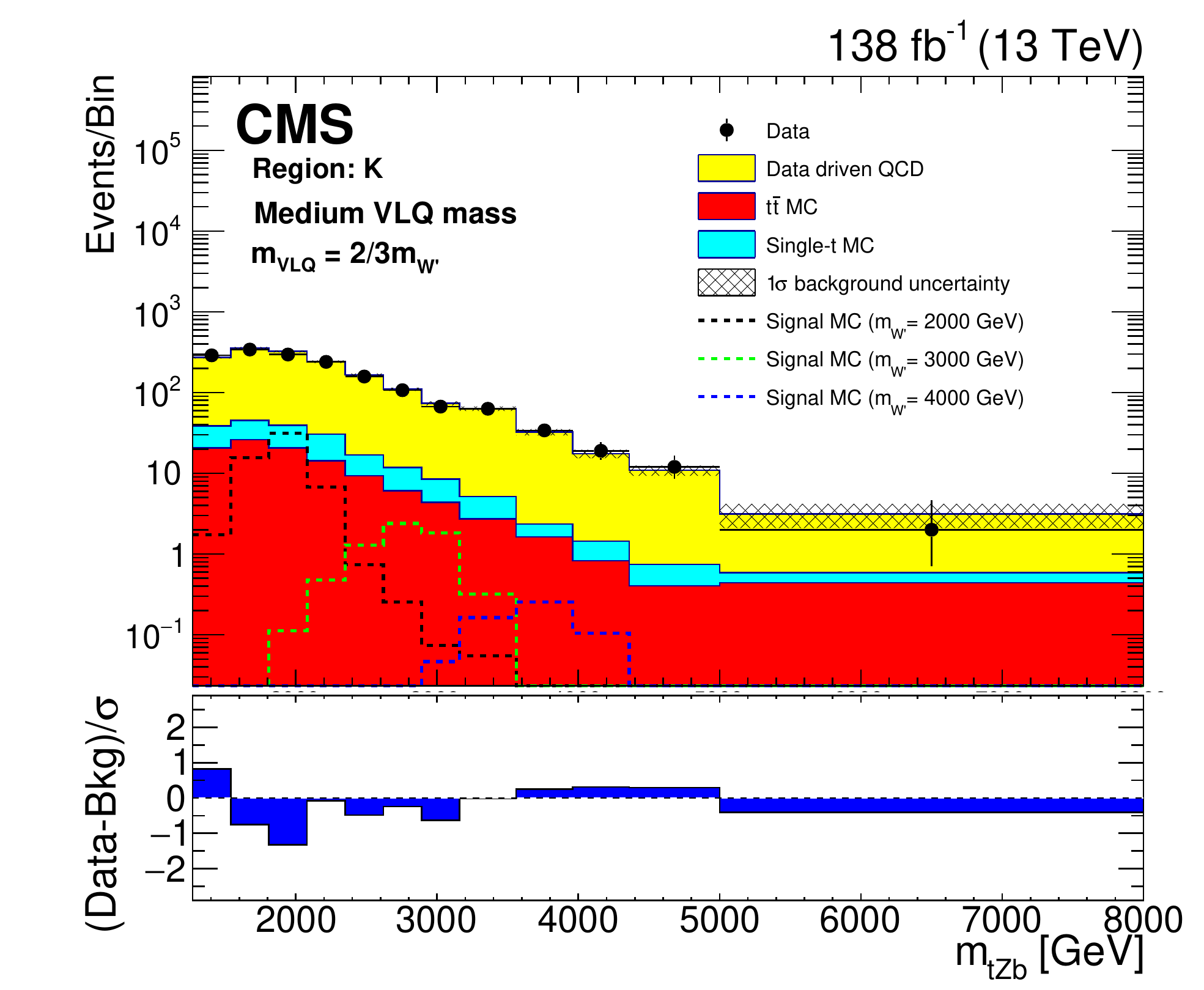}\\
\includegraphics[width=0.45\textwidth]{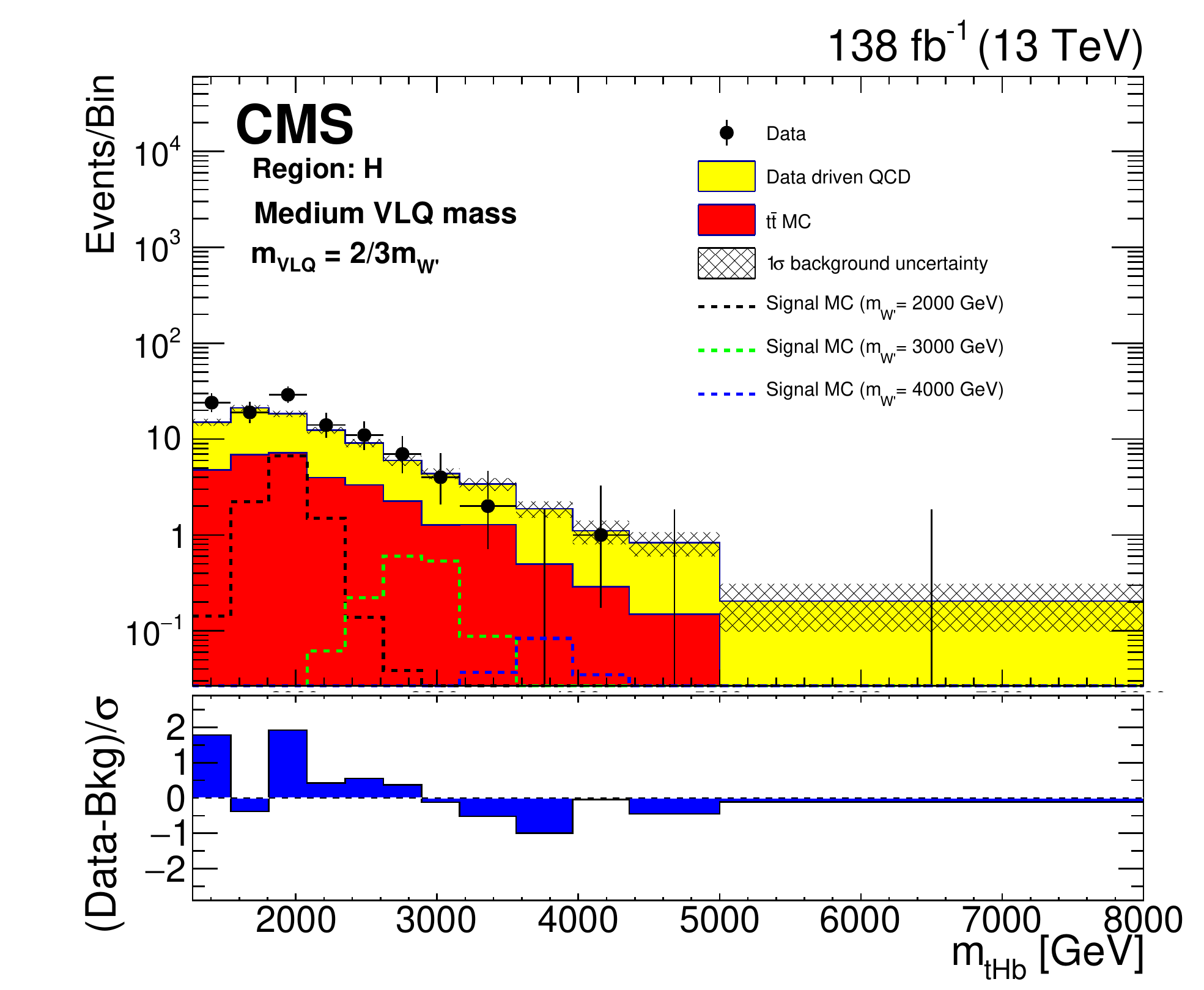}
\includegraphics[width=0.45\textwidth]{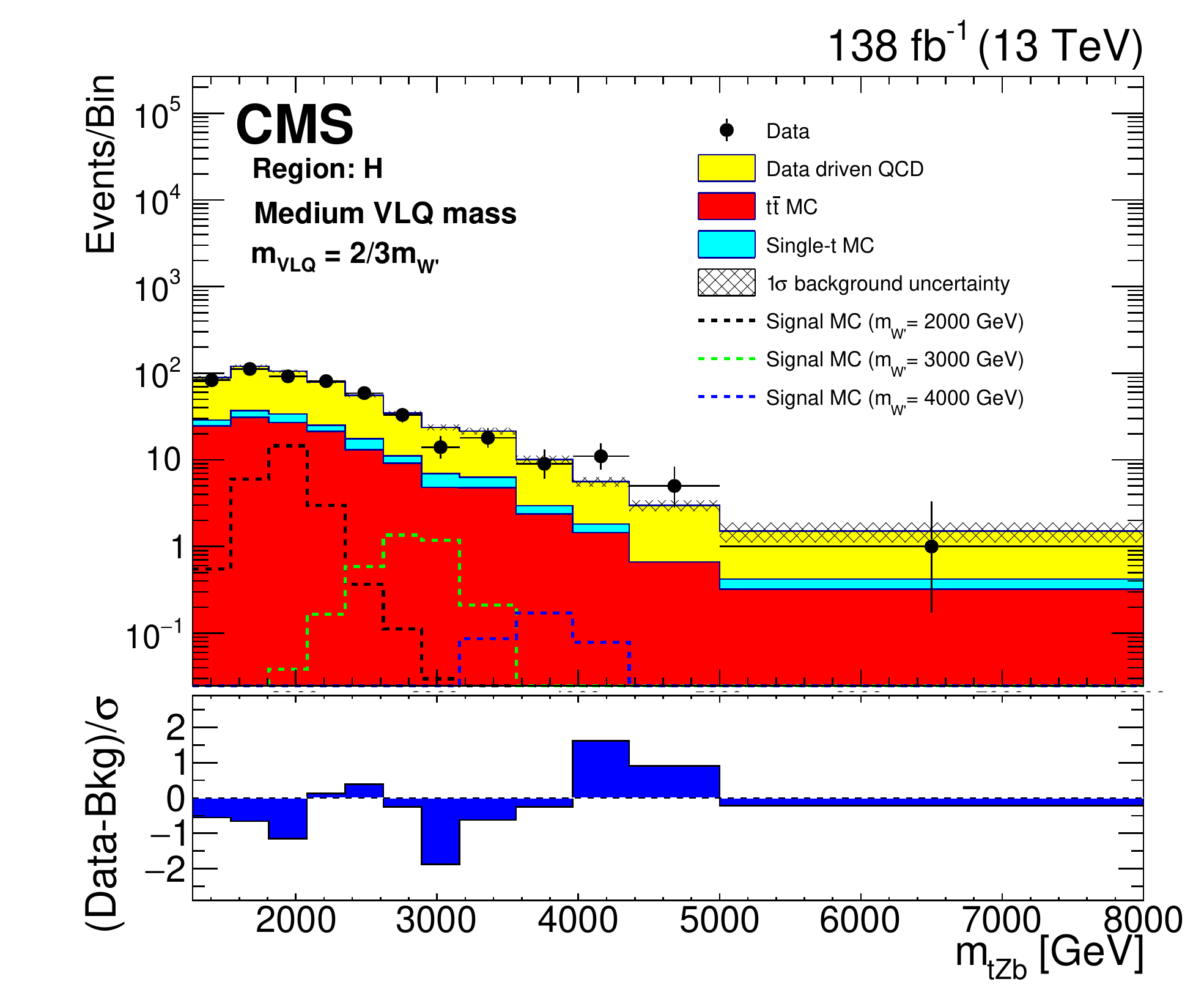}\\
\caption{Background closure test for the reconstructed \PWpr boson invariant mass
in region F (upper), K (middle), and H (lower) for the
purpose of validation in the \tHb (left) and \tZb (right) analyses.
The data are shown as points with error bars and the estimated backgrounds
as solid histograms after a background-only fit, with a hatched band indicating the uncertainty.
Expected signals for several \PWpr boson mass hypotheses are shown.
The lower panel of each plot shows the
difference between the number of events observed in the data and the predicted
background, divided by the total uncertainty in the background and the
statistical uncertainty in the data added in quadrature.
}
\label{figs:invmthbval201620172018}
\end{figure}

To test the validity of the background estimate in
data, a series of ``validation'' regions are defined (F, K, and H, as defined in Fig.~\ref{figs:cartoon}).  The
transfer function used for predicting the background in the K and F regions (\TFV) is estimated from the ratio
of regions E to A, whereas the transfer function for the H region is the same as for the signal region.
The following relations are used to extract the QCD background component of the validation regions:
\begin{linenomath}
\begin{equation}
\label{equation:regionsval}
\begin{split}
\TFV &\equiv (\mathrm{E}_{\text{data}}-\mathrm{E}_{\ttbar})/(\mathrm{A}_{\text{data}}-\mathrm{A}_{\ttbar}),\\
\mathrm{H}_{\text{QCD}} &\simeq (\mathrm{G}_{\text{data}}-\mathrm{G}_{\ttbar}) \TF, \\
\mathrm{K}_{\text{QCD}} &\simeq (\mathrm{D}_{\text{data}}-\mathrm{D}_{\ttbar}) \TFV,\\
\mathrm{F}_{\text{QCD}} &\simeq (\mathrm{G}_{\text{data}}-\mathrm{G}_{\ttbar}) \TFV.
\end{split}
\end{equation}
\end{linenomath}
The background validation tests in the F, K, and H regions are shown in Fig.~\ref{figs:invmthbval201620172018}.
These validation regions confirm the background estimate agreement
with $\chi^2$/ndf values of 1.7, 1.0, and 1.3, respectively for the combined fit which
considers systematic uncertainties (as discussed in Section~\ref{sec:systematics}), where ndf is the number of degrees of freedom.
The \ttbar background component in these validation regions is subtracted using the same procedure
that is used when performing the signal region background estimate.  The agreement demonstrates
that the \PQt jet selection can be inverted without biasing the Higgs jet selection.
Figure~\ref{figs:invmthbval201620172018} also shows for comparison the distributions expected from a potential signal for
several \PWpr mass values, assuming the medium VLQ mass hypothesis. The corresponding distributions
for the low and high VLQ mass values have shapes that are similar to these, differing mainly in the normalization.

The \ttbar background component is estimated using simulated events with an additional event
weight that corrects the generator \PQt jet \pt distribution~\cite{PhysRevD.95.092001}.
This correction reduces the disagreement between simulation and data
due to a known generator-level mismodeling effect and is validated
using a \ttbar control region.  The final state for this control region contains
two boosted AK8 \PQt jets, so the selections on the AK4 \PQb jet are loosened, and
both of the AK8 jets are required to pass the same \mdtop selection as the signal region.
After this selection, the composition is a nearly pure (approximately 99\%) sample of hadronic \ttbar events,
demonstrating the discrimination power of the \mdtop tagger.  The QCD
background estimate in this control region is performed in an analogous way to the signal region, and
is estimated by weighting events with one inverted and one tight \PQt tag by \TF.
This control region is used to extract an additional correction, which
is taken as a simple ratio of QCD-subtracted data to the
\ttbar simulation estimate.  The uncertainty in this procedure is taken as
the uncorrelated combination of the statistical uncertainty in the
ratio, 100\% of the QCD estimate, and the \PQt tag scale factor uncertainty.
This correction is applied to all simulated \ttbar distributions.

\section{Systematic uncertainties}
\label{sec:systematics}
A wide range of systematic uncertainties are considered, and
organized into those that impact only the event yields, which
are assumed to follow log-normal distributions~\cite{Conway:2011in},
and those that affect the shape of the three-jet invariant mass distribution as well,
which are assumed to follow Gaussian distributions. All of the systematic uncertainties are summarized in
Table~\ref{table:syst}.

\begin{table}[htb]
\topcaption{Sources of systematic uncertainty that affect the final distributions.
Sources that affect the normalization only are represented by a range of values for the systematic variation.
Sources listing the systematic variation as ${\pm}1\sigma(x)$ affect the shape of
the three-jet invariant mass distribution as well, which is dependent on $x$.}
\label{table:syst}
\centering
\begin{tabular}{lcc}
{Source} & {Variation} & {Process}  \\
\hline
Integrated luminosity & $\pm$1.2--2.5\% & signal, \ttbar, \st \\
\ttbar normalization & $\pm$11--15\% & \ttbar  \\
JES & $\pm 1\sigma (\pt, \eta)$ & signal, \ttbar, \st \\
JER & $\pm 1\sigma (\pt, \eta)$ & signal, \ttbar, \st \\
JMS & $\pm 1\sigma (m_{\mathrm{SD}})$ & signal, \ttbar, \st \\
JMR & $\pm 1\sigma (m_{\mathrm{SD}})$ & signal, \ttbar, \st\\
\PZ jet tagging & $\pm 1\sigma (\pt)$ & signal\\
\PQb  tagging & $\pm 1\sigma (\pt)$ &  signal, \ttbar, \st \\
\PQb  mistagging & $\pm 1\sigma (\pt)$ &  signal, \ttbar, \st \\
\db tagging & $\pm 1\sigma (\pt)$ & signal\\
\db mistagging & $\pm 1\sigma (\pt)$ & signal, \ttbar, \st  \\
\PQt jet tagging &  $\pm 1\sigma (\pt)$ &  signal, \ttbar, \st \\
Trigger & $\pm 1\sigma (\HT)$ & signal, \ttbar, \st  \\
Pileup & $\pm 1\sigma$ & signal, \ttbar, \st \\
PDF, $\alpha_{s}$ & $\pm 1\sigma$ &  signal, \st  \\
$\mu_\mathrm{R}$ and $\mu_\mathrm{F}$ & $\pm 1\sigma$ &  signal, \st \\
ISR/FSR & $\pm 1\sigma$ &  \st \\
\TF & $\pm 1\sigma (\pt, \eta)$ & QCD \\
\ttbar contamination & $\pm 1\sigma (\pt, \eta)$ & QCD \\
\end{tabular}
\end{table}

\subsection{Normalization uncertainties}

The uncertainty in the integrated luminosity is taken as 1.2\% for 2016, 2.3\%
for 2017, and 2.5\% for 2018 data~\cite{CMS:2021xjt,CMS-PAS-LUM-17-004,CMS-PAS-LUM-18-002}.

The uncertainty in the \ttbar normalization measurement (as discussed in Section~\ref{sec:backgroundestimation})
used to correct the estimate in the signal region accounts for the full uncertainty in the \ttbar yield.  It is
calculated separately for data collected in 2016, 2017, and 2018, and ranges from
11 to 15\%.  This is considered a measurement of \ttbar uncertainties that have a large normalization effect,
including top \pt reweighting, PDF, and renormalization and factorization
($\mu_\mathrm{R}$ and $\mu_\mathrm{F}$) scale uncertainties.
Uncertainties that have a pronounced shape effect on the invariant mass
distribution are included separately when setting limits (as shown in Table~\ref{table:syst}).

\subsection{Shape uncertainties}

The uncertainty in the JES is taken into account by varying the JES within
its uncertainty and observing the effect on the invariant mass distribution.
The JES variation impacts the invariant mass distribution shape through a
horizontal shift but also causes a normalization difference when
the jet falls above or below a kinematic threshold.  This uncertainty is
propagated to the \PQt jet \msdt estimate.  The uncertainty in the
JER is also taken into account
by the uncertainty in the JER correction used
for simulated events.  This uncertainty is applied to all simulated events and has only a small impact on the final limit.

The uncertainty in the JMS and JMR is measured in a highly
enriched sample of \ttbar events containing one final-state lepton.  In this sample,
a fit is performed to the \PW boson PUPPI \msd distribution, from which the
mean and width are extracted.  The JMS uncertainty is
estimated from the uncertainty in the shift of the \PW mass peak, and the uncertainty
in the JMR is estimated from the uncertainty in the width.
These uncertainties are applied to the Higgs or \PZ boson \msd estimates for all simulated
events and have an $\approx$4\% effect on the signal estimate.

The uncertainty in the \ztau selection associated with \PZ jet tagging includes a constant uncertainty
and a \pt-dependent uncertainty
based on an extrapolation to momenta beyond the kinematic region of the scale factor
measurement.  This uncertainty is applied to all simulated events
and has an $\approx$15\% effect.

The uncertainty used for the \PQb tagging requirement on the AK4 jet is
evaluated by varying the \PQb tagging and \PQb mistagging scale factors
within their uncertainties~\cite{Sirunyan:2017ezt}.  These uncertainty
sources are considered uncorrelated and are applied to all simulated events and
result in an $\approx$5\% effect on the signal estimate.

The uncertainty in the \db method used for the Higgs jet tagging~\cite{Sirunyan:2017ezt}
selection is evaluated by varying the \db scale factor by the uncertainty.
The scale factor is parameterized using three regions in \pt.
Also evaluated is the scale factor for a Higgs boson that is
mistagged as a \PQt quark (as discussed in Section~\ref{sec:eventreconstruction}).
The uncertainties in both the Higgs jet tagging efficiency and the mistag probability are
applied to all simulated events and are treated as
uncorrelated during limit setting.  The Higgs boson \db uncertainty
has an $\approx$10\% effect on the signal estimate.

The uncertainty in the \mdtop selection used for \PQt tagging is \pt dependent
and evaluated by varying the \mdtop scale factor within the uncertainty.
This has an $\approx$6\% effect on the signal estimate.

The kinematic requirements ensure that the trigger
is close to being fully efficient, and all simulated events are
corrected using the method outlined in Section~\ref{sec:eventreconstruction}.
The uncertainty in this correction is taken to be half of the trigger inefficiency.  This uncertainty
is small and is applied to all simulated events.

As mentioned in Section~\ref{sec:datasample}, the simulated pileup distribution
is reweighted to match data using an effective total inelastic cross section.
The uncertainty in this procedure is evaluated by varying the
total inelastic cross section by $\pm$4.6\%~\cite{Sirunyan:2018nqx}.  The resultant
uncertainty is applied to all simulated events and has
only a small impact on the final limit.

The PDF uncertainty is evaluated using the NNPDF3.1~\cite{Ball:2014uwa,Ball:2017nwa}
set for the signal, and the PDF4LHC15 combined set~\cite{Botje:2011sn,Alekhin:2011sk}
for single \PQt quark events.
The uncertainty is calculated using the symmetric Hessian approach.

The uncertainty in the $\mu_\mathrm{R}$ and $\mu_\mathrm{F}$ scales are evaluated
by varying the
$\mu_\mathrm{R}$ or $\mu_\mathrm{F}$ scales up and down by a factor of two.
Both independent and simultaneous
variations of $\mu_\mathrm{R}$ and $\mu_\mathrm{F}$ are considered, resulting in six
weights after excluding the two most extreme variations.  The envelope of these weights is
used as the uncertainty.
The $\mu_\mathrm{R}$, $\mu_\mathrm{F}$, and PDF uncertainties are
taken as the signal cross section
theoretical uncertainty.  For the single \PQt quark prediction, these uncertainties are
considered when setting limits in addition to the parton shower variation in the
initial- and final-state radiation (ISR and FSR).
Together, the effect of these theoretical uncertainties amount to $\approx$65\%,
and dominate the total uncertainty in the single top quark estimate.

The primary uncertainty in the shape of the QCD
background estimate is taken from the statistical uncertainty in the \TF.  This is
propagated to the invariant mass spectrum by evaluating the \TF weight at ${\pm}1\sigma$
in a given (\pt, $\eta$) bin.  The uncertainty from each \TF bin is added in
quadrature to form the full uncertainty, and the effect on the QCD background estimate is $\approx$20\%.
The effect of the uncertainty in the \ttbar subtraction procedure is
evaluated by varying the amount subtracted within the range of the \ttbar normalization
uncertainty. The effect on the QCD background estimate is $\approx$6\%.

The statistical uncertainty of the simulation is taken into account during limit setting by using the
``Barlow--Beeston lite'' method~\cite{BARLOW1993219}.

\section{Results}
\label{sec:limits}

\begin{figure}[htbp]
\centering
\includegraphics[width=0.75\textwidth]{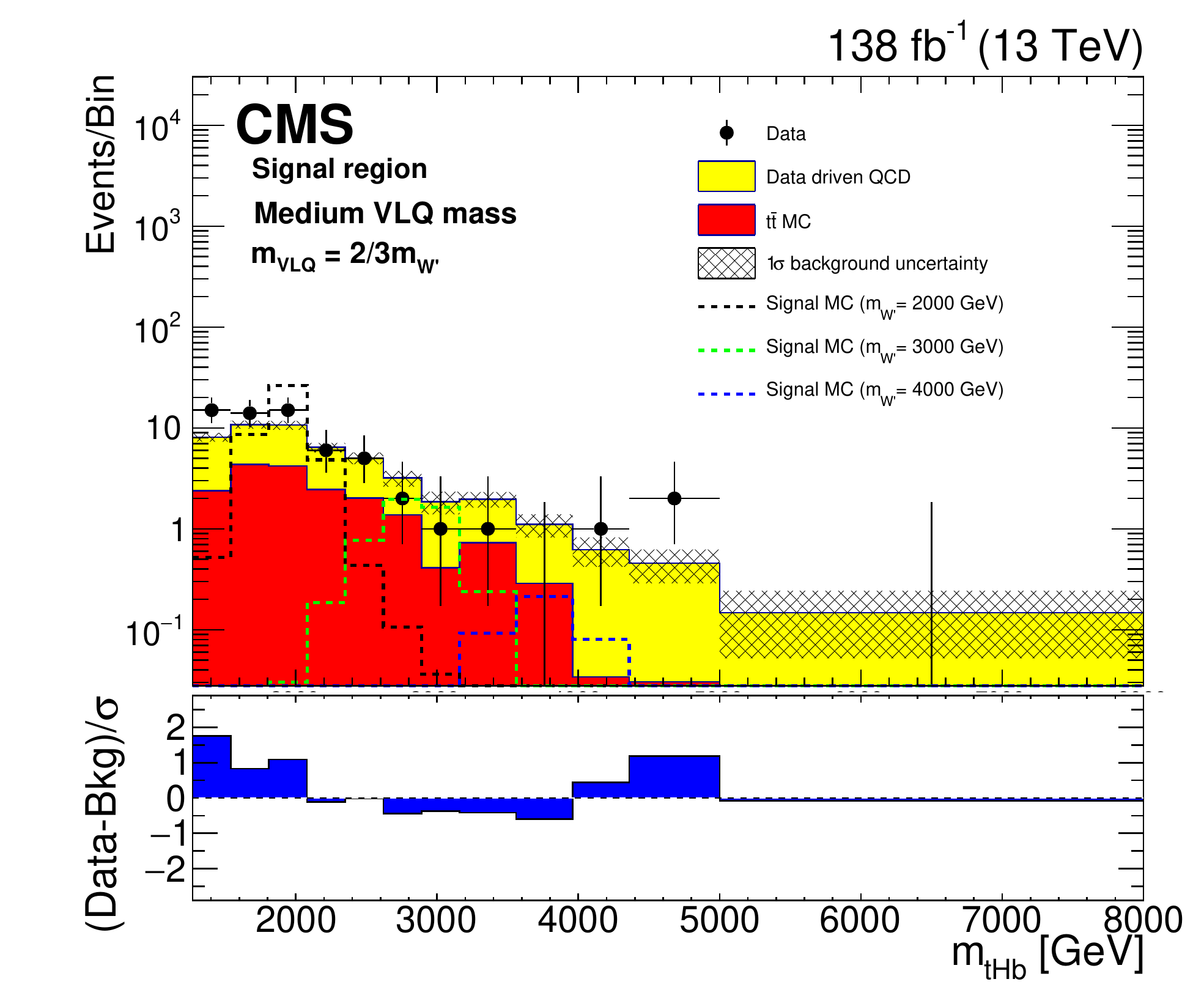}\\
\includegraphics[width=0.75\textwidth]{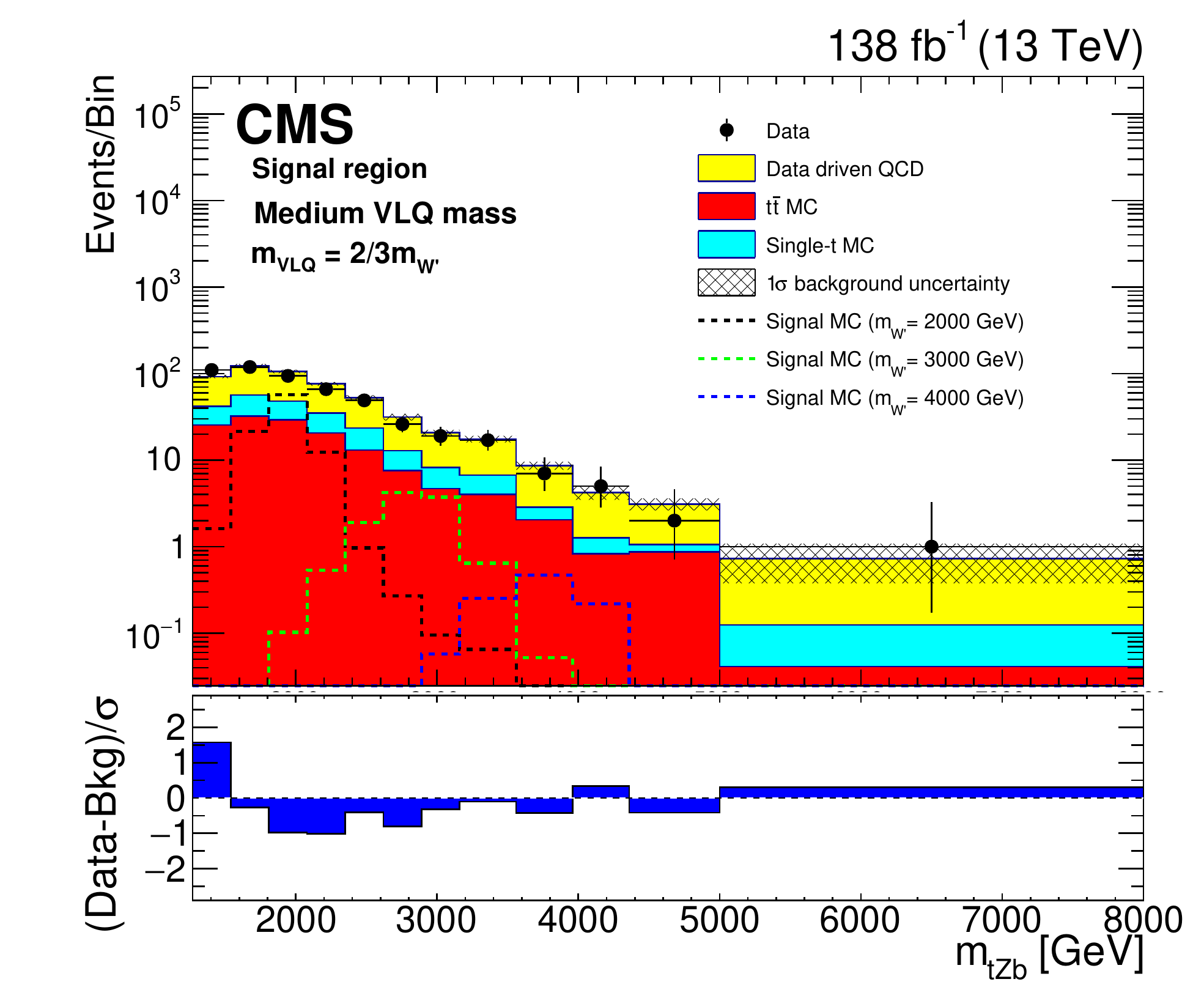}
\caption{Reconstructed $m_{\PWpr}$ distributions (\mthb (upper), and \mtzb
(lower)) in the signal region with estimated backgrounds and signal for several \PWpr boson
mass hypotheses, after a background-only fit. The combined statistical and systematic uncertainty in the
total estimated background is indicated by the hatched region.  The lower panels show the
difference between the number of events observed in the data and the predicted
background, divided by the total uncertainty in the background and the
statistical uncertainty in the data added in quadrature.
}
\label{figs:SRwpmass}
\end{figure}

The final distributions of reconstructed $m_{\PWpr}$ for the signal region
are shown in Fig.~\ref{figs:SRwpmass}.
The distributions expected from a potential signal for several \PWpr mass values,
assuming the medium VLQ mass hypothesis, are shown for comparison.
Since no deviation from the SM expectation is observed, we place limits on the \PWpr
boson production cross section.

\begin{figure}[htbp]
\centering
\includegraphics[width=0.6\textwidth]{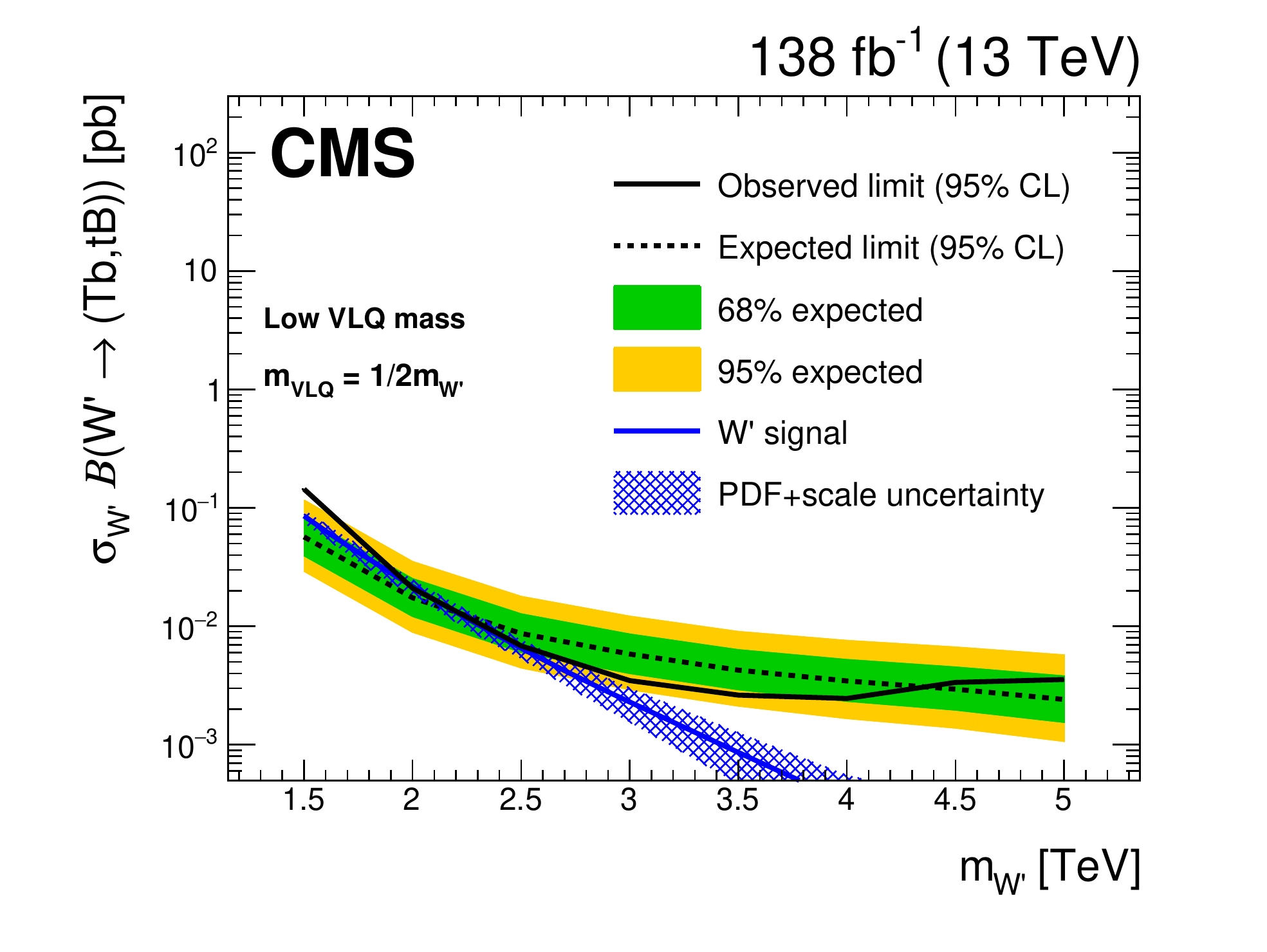}\\
\includegraphics[width=0.6\textwidth]{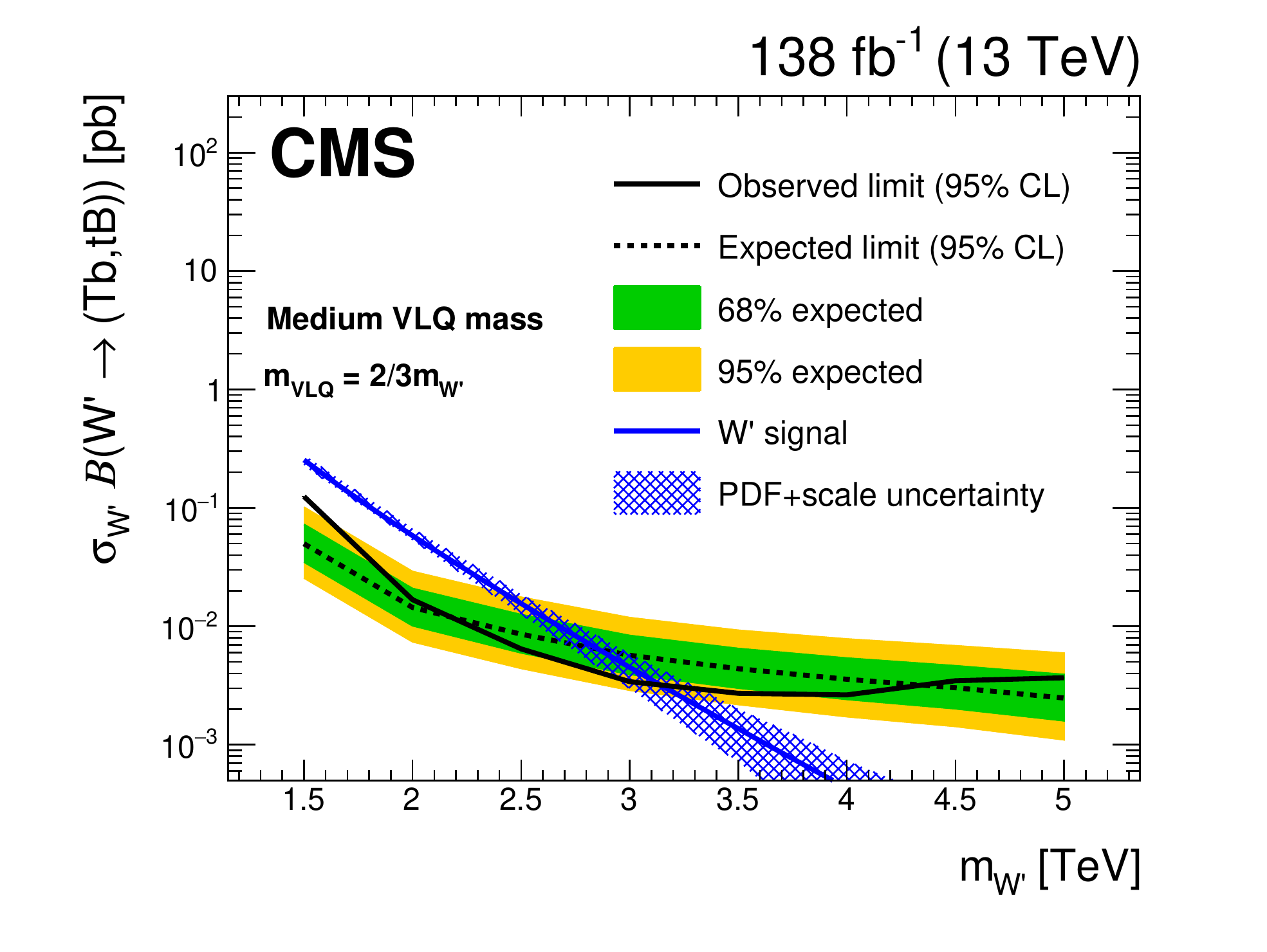}\\
\includegraphics[width=0.6\textwidth]{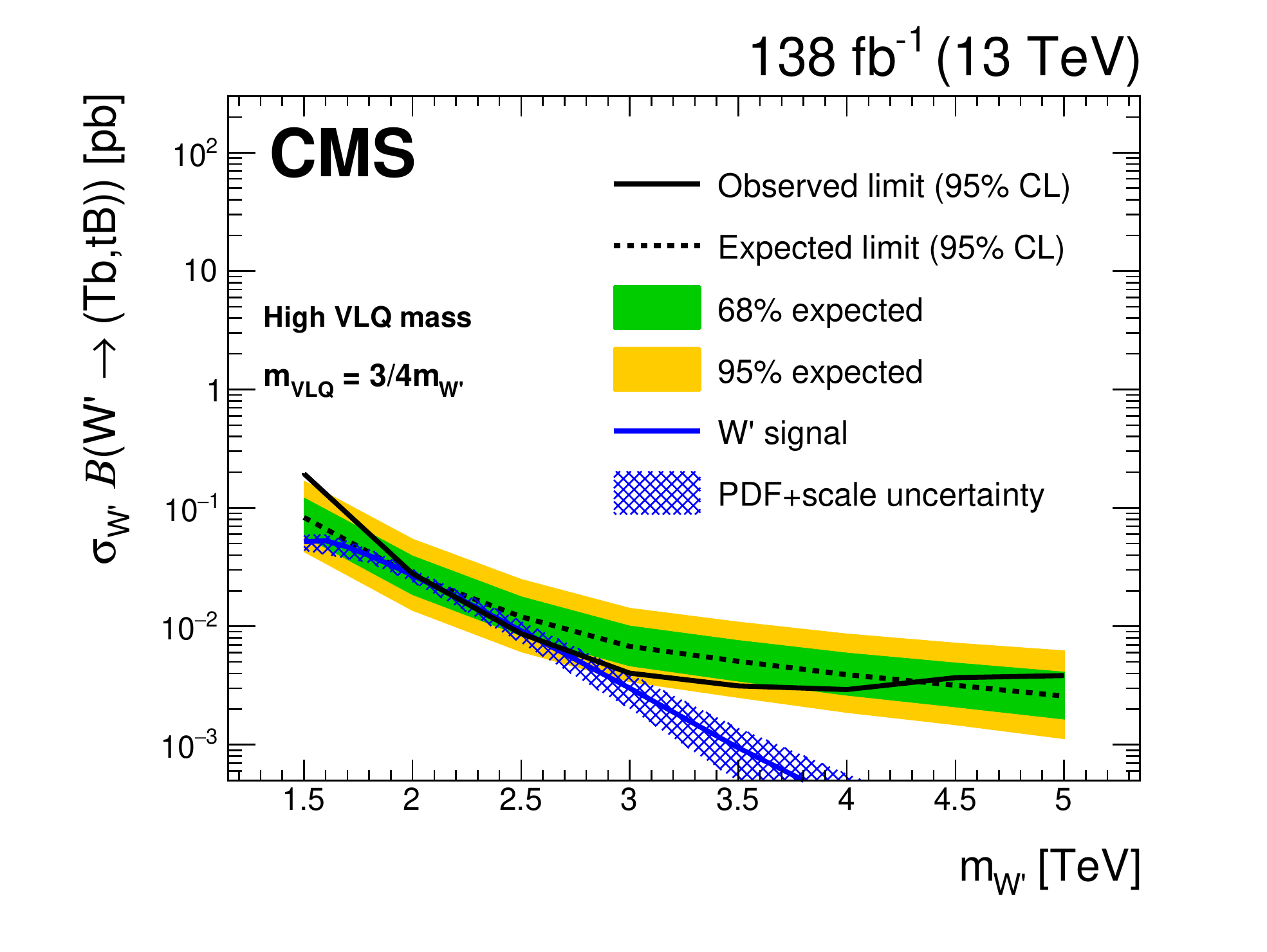}\\
\caption{
The \PWpr boson 95\% \CL limits on the product of cross section and branching fraction.
The expected (dashed) and observed (solid) limits, as well as the \PWpr boson
theoretical cross section, with its PDF and scale normalization uncertainties, are shown.
The green (inner) and yellow (outer) bands indicate the 68\% and 95\% confidence intervals of the expected limit.
The limits are given for the low (upper),
medium (center), and high (lower) VLQ mass ranges.
}
\label{figs:limits}
\end{figure}

The 95\% confidence level (\CL) upper
limits are obtained using the \CLs approach~\cite{Junk:1999kv,Read:2002hq}
in the asymptotic approximation~\cite{Cowan:2010js} on the product of the \PWpr
boson production cross section
for the $s_{\mathrm{L}}=0.5$ and $\cott=3$ hypothesis~\cite{Vignaroli:2014bpa},
and for the benchmark branching fraction.
A binned maximum likelihood fit to the data is used to calculate upper limits, in a process where
all systematic uncertainties are included as nuisance parameters.  For the signal template, the sum
of the invariant mass distributions from the $\PQt\bpr$ and $\PQb\tpr$ decay
channels are used.  Upper limits on the product of cross section and branching fraction are shown in Fig.~\ref{figs:limits}.
A \PWpr boson with a mass below 3.1\TeV is excluded at 95\%~\CL for the medium
VLQ mass case.  The low and high VLQ mass benchmarks have a lower $\PWpr \to \text{VLQ}$ branching fraction, and
the sensitivity is not sufficient to set mass exclusion limits.

\begin{figure}[htbp]
\centering
\includegraphics[width=0.45\textwidth]{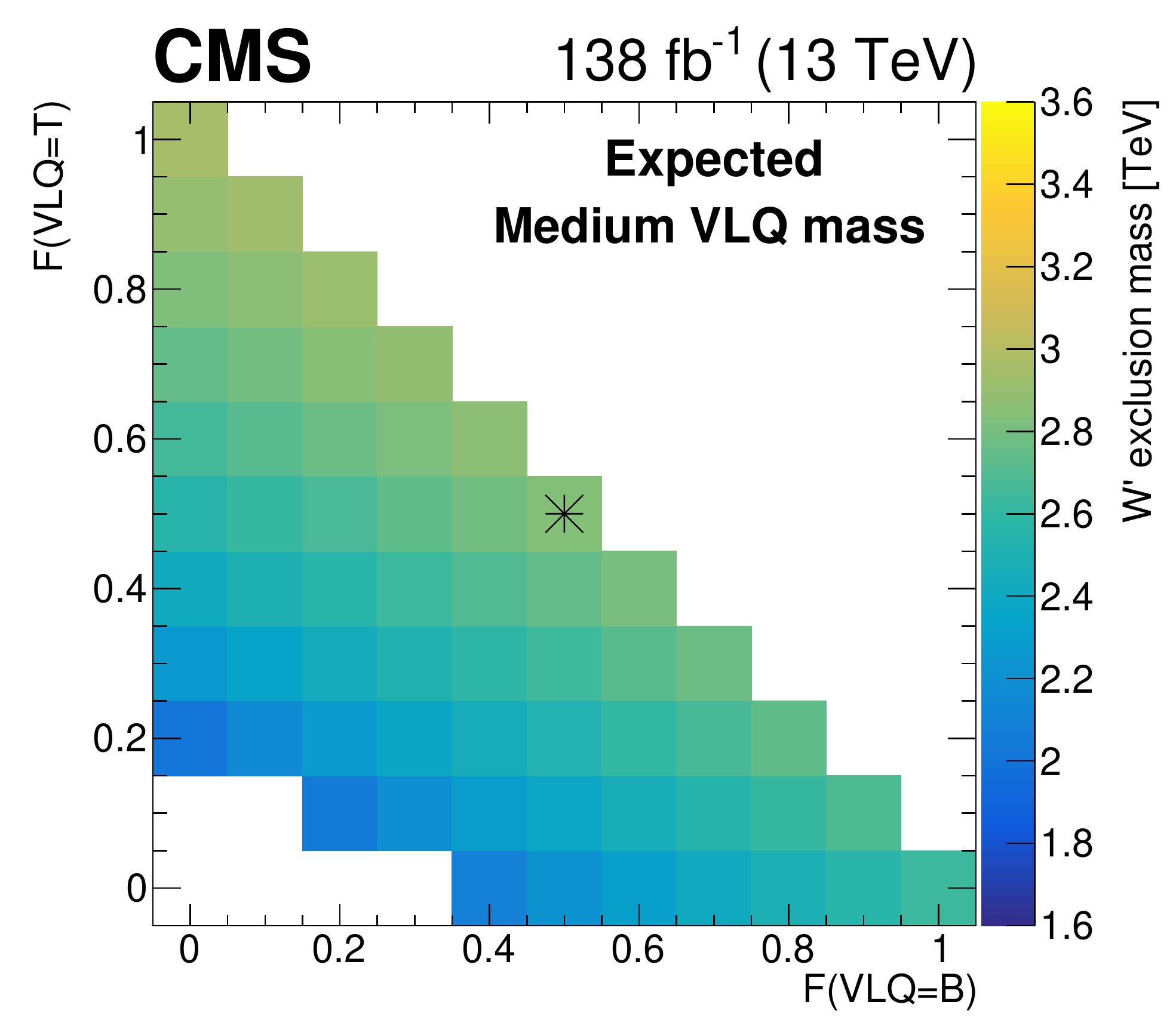}
\includegraphics[width=0.45\textwidth]{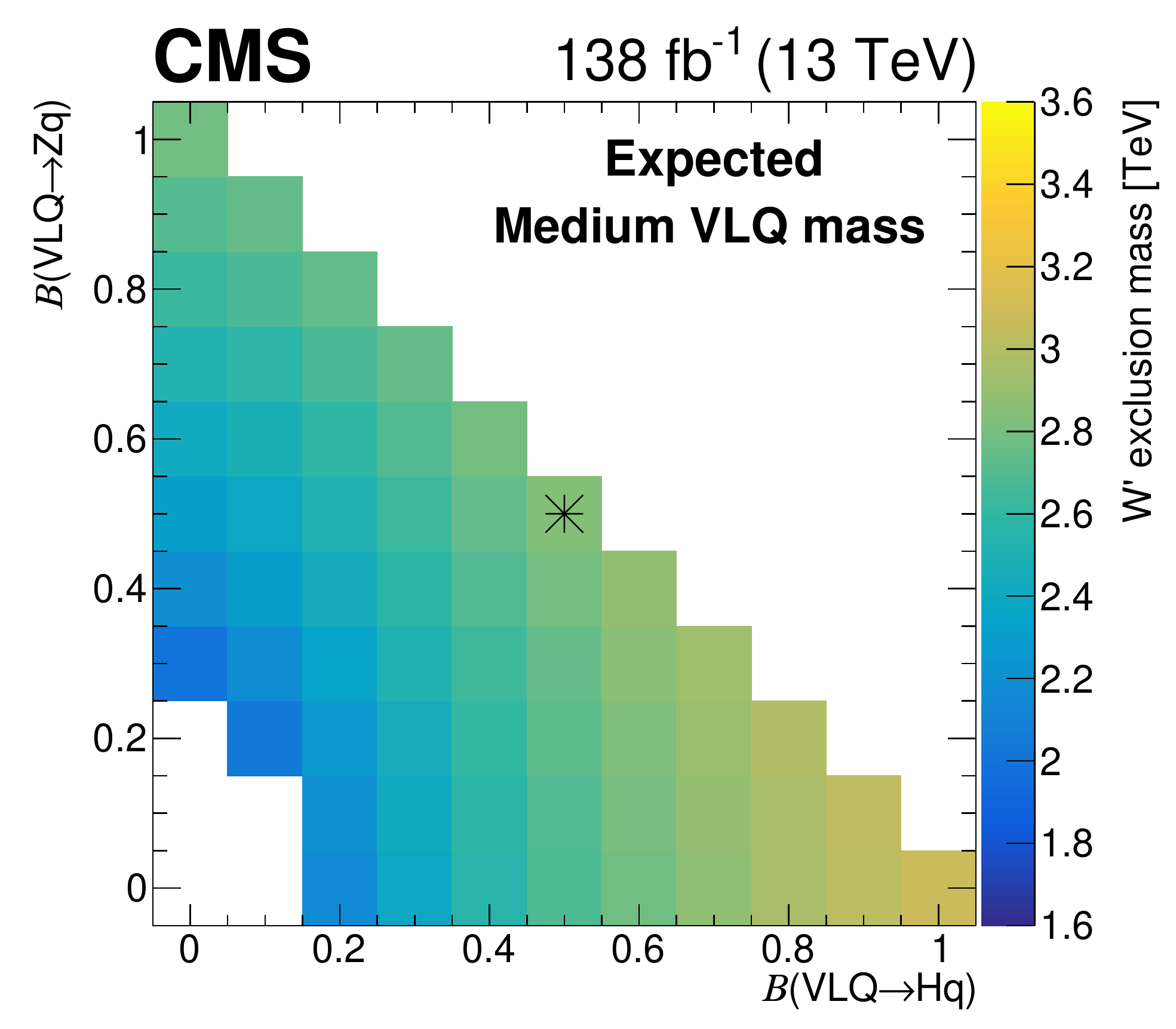}\\
\includegraphics[width=0.45\textwidth]{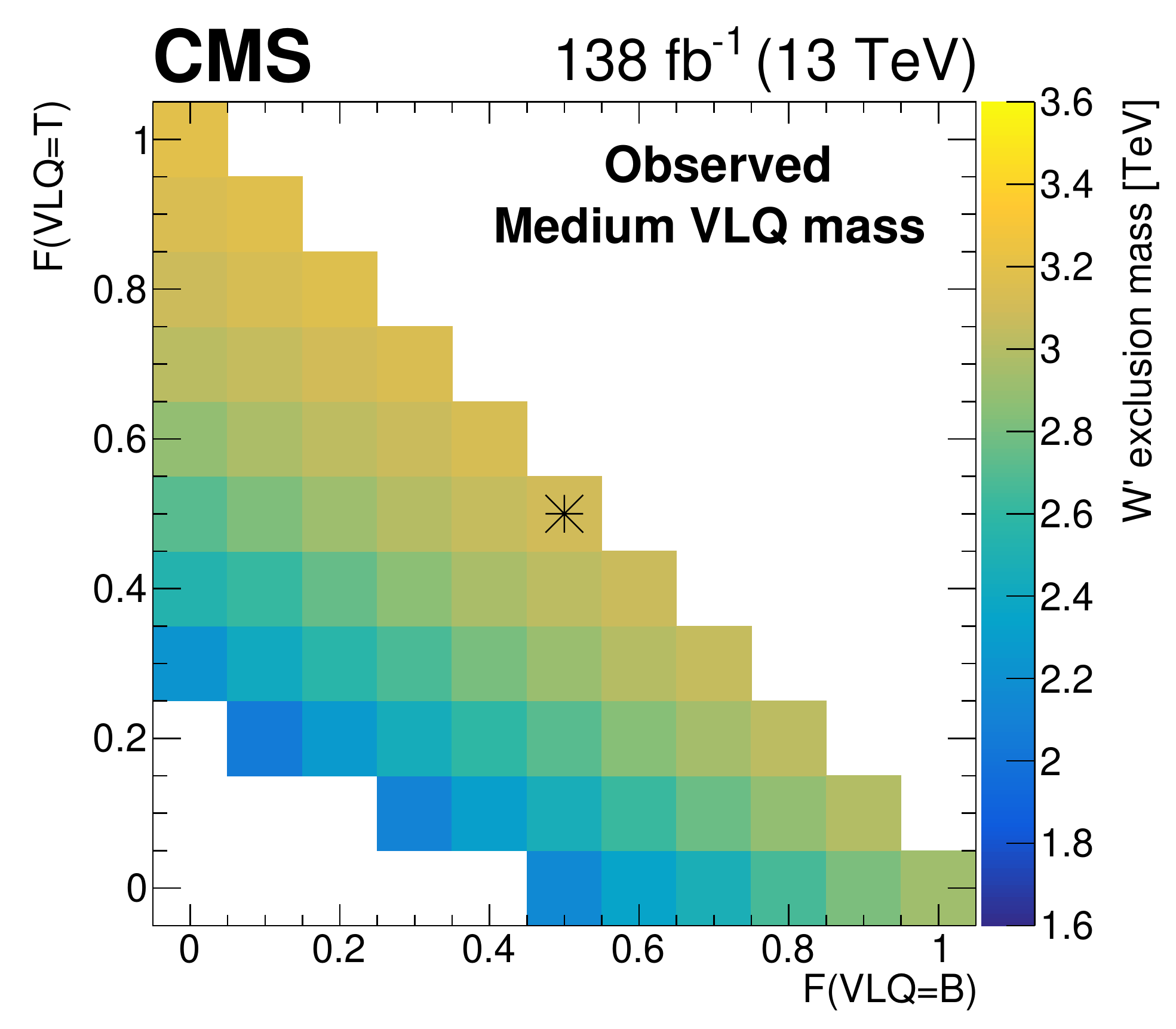}
\includegraphics[width=0.45\textwidth]{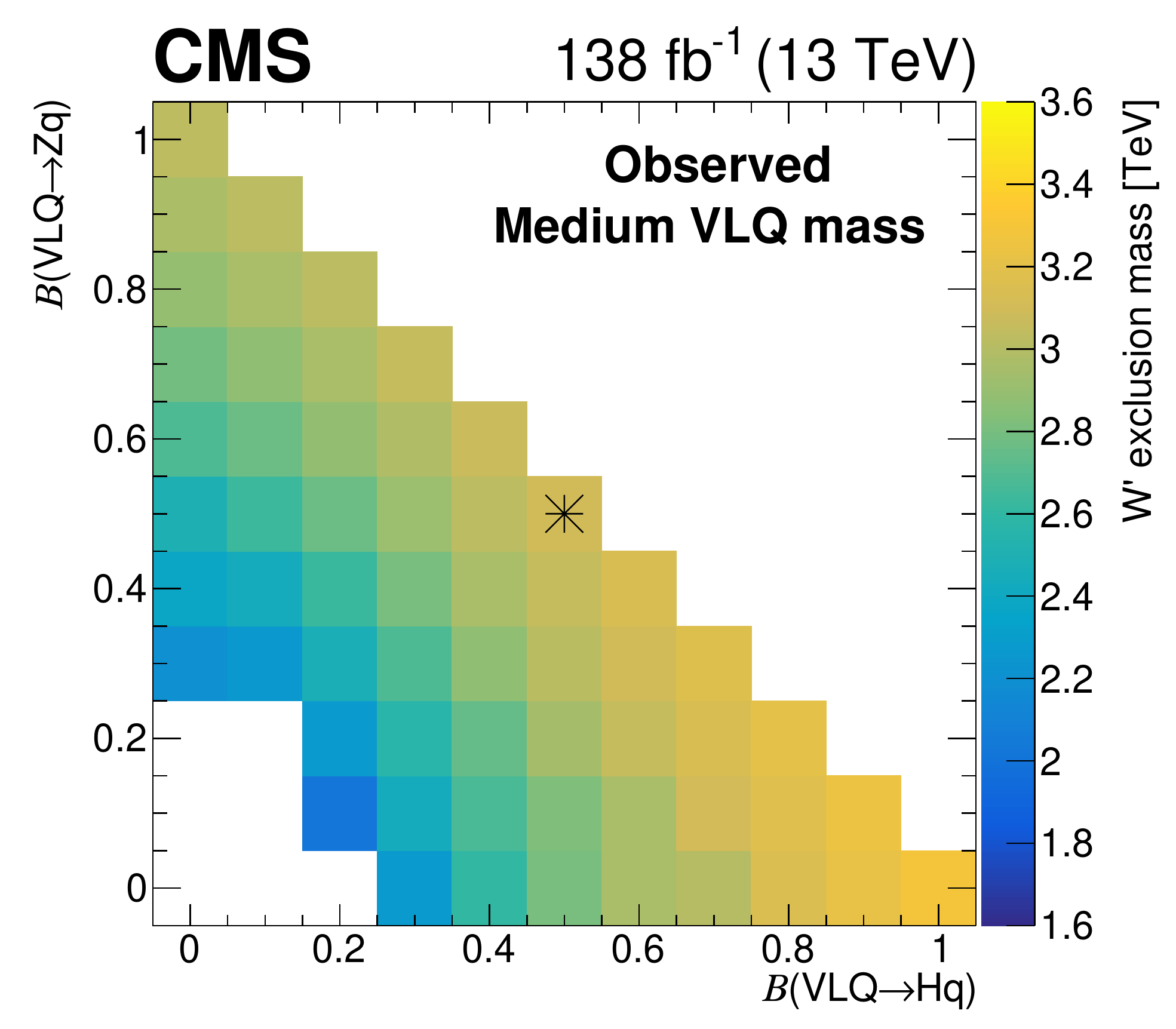}\\
\caption{Expected (upper) and observed (lower) 95\% \CL limits for generalized hypotheses varying the
fraction of $\PQt\bpr$ (F(VLQ=\bpr)) and $\PQb\tpr$ (F(VLQ=\tpr)) from the \PWpr decay (left), and the VLQ branching fraction to $\PQq\PH$ and $\PQq\PZ$  (right).
The asterisk marker signifies the branching fractions for the benchmark model
}
\label{figs:genlimits}
\end{figure}

Variations of the primary assumptions of this benchmark are shown in
Fig.~\ref{figs:genlimits} for the purpose of generalizing the results.  Specifically, the
fraction of $\PQb\tpr$ and $\PQt\bpr$ from the \PWpr decay is varied between 0 and
1, and the VLQ branching fractions to $\PQq\PH$ and $\PQq\PZ$ are varied within the
same bounds.  Therefore, for both generalized coupling figures, the mass exclusion
of the benchmark model corresponds to the point with coordinates $[0.5, 0.5]$, indicated with an asterisk.

\section{Summary}
A search has been presented for a heavy \PWpr boson decaying to a \bpr or \tpr vector-like
quark and a \PQt or \PQb quark, respectively. The data
correspond to an integrated luminosity of 138\fbinv collected between 2016 and 2018
with the CMS detector at the LHC in proton-proton collisions at $\sqrt{s}=13\TeV$. The signature considered for both decay modes
is a \PQt quark and a Higgs or \PZ boson, each decaying hadronically, and a \PQb quark jet.
Boosted heavy-resonance identification techniques are used to select the events containing
three energetic jets and to suppress standard model backgrounds.
No significant deviation from the standard model background prediction is
observed.  Upper limits are placed on the product of
the \PWpr boson cross section and the final state branching fraction as a
function of the $m_{\PWpr}$.
A \PWpr boson with a mass below 3.1\TeV is excluded at 95\% confidence level, given the benchmark model
assumption of democratic branching fractions.
In addition, limits are set based on generalizations of these assumptions.
These are the most sensitive limits to date for this final state.

\begin{acknowledgments}
	We congratulate our colleagues in the CERN accelerator departments for the excellent performance of the LHC and thank the technical and administrative staffs at CERN and at other CMS institutes for their contributions to the success of the CMS effort. In addition, we gratefully acknowledge the computing  centers and personnel of the Worldwide LHC Computing Grid and other  centers for delivering so effectively the computing infrastructure essential to our analyses. Finally, we acknowledge the enduring support for the construction and operation of the LHC, the CMS detector, and the supporting computing infrastructure provided by the following funding agencies: BMBWF and FWF (Austria); FNRS and FWO (Belgium); CNPq, CAPES, FAPERJ, FAPERGS, and FAPESP (Brazil); MES and BNSF (Bulgaria); CERN; CAS, MoST, and NSFC (China); MINCIENCIAS (Colombia); MSES and CSF (Croatia); RIF (Cyprus); SENESCYT (Ecuador); MoER, ERC PUT and ERDF (Estonia); Academy of Finland, MEC, and HIP (Finland); CEA and CNRS/IN2P3 (France); BMBF, DFG, and HGF (Germany); GSRI (Greece); NKFIA (Hungary); DAE and DST (India); IPM (Iran); SFI (Ireland); INFN (Italy); MSIP and NRF (Republic of Korea); MES (Latvia); LAS (Lithuania); MOE and UM (Malaysia); BUAP, CINVESTAV, CONACYT, LNS, SEP, and UASLP-FAI (Mexico); MOS (Montenegro); MBIE (New Zealand); PAEC (Pakistan); MSHE and NSC (Poland); FCT (Portugal); JINR (Dubna); MON, RosAtom, RAS, RFBR, and NRC KI (Russia); MESTD (Serbia); MCIN/AEI and PCTI (Spain); MOSTR (Sri Lanka); Swiss Funding Agencies (Switzerland); MST (Taipei); ThEPCenter, IPST, STAR, and NSTDA (Thailand); TUBITAK and TAEK (Turkey); NASU (Ukraine); STFC (United Kingdom); DOE and NSF (USA).

	\hyphenation{Rachada-pisek} Individuals have received support from the Marie-Curie program and the European Research Council and Horizon 2020 Grant, contract Nos.\ 675440, 724704, 752730, 758316, 765710, 824093, 884104, and COST Action CA16108 (European Union); the Leventis Foundation; the Alfred P.\ Sloan Foundation; the Alexander von Humboldt Foundation; the Belgian Federal Science Policy Office; the Fonds pour la Formation \`a la Recherche dans l'Industrie et dans l'Agriculture (FRIA-Belgium); the Agentschap voor Innovatie door Wetenschap en Technologie (IWT-Belgium); the F.R.S.-FNRS and FWO (Belgium) under the ``Excellence of Science -- EOS" -- be.h project n.\ 30820817; the Beijing Municipal Science \& Technology Commission, No. Z191100007219010; the Ministry of Education, Youth and Sports (MEYS) of the Czech Republic; the Deutsche Forschungsgemeinschaft (DFG), under Germany's Excellence Strategy -- EXC 2121 ``Quantum Universe" -- 390833306, and under project number 400140256 - GRK2497; the Lend\"ulet (``Momentum") Program and the J\'anos Bolyai Research Scholarship of the Hungarian Academy of Sciences, the New National Excellence Program \'UNKP, the NKFIA research grants 123842, 123959, 124845, 124850, 125105, 128713, 128786, and 129058 (Hungary); the Council of Science and Industrial Research, India; the Latvian Council of Science; the Ministry of Science and Higher Education and the National Science Center, contracts Opus 2014/15/B/ST2/03998 and 2015/19/B/ST2/02861 (Poland); the Funda\c{c}\~ao para a Ci\^encia e a Tecnologia, grant CEECIND/01334/2018 (Portugal); the National Priorities Research Program by Qatar National Research Fund; the Ministry of Science and Higher Education, projects no. 0723-2020-0041 and no. FSWW-2020-0008 (Russia); MCIN/AEI/10.13039/501100011033, ERDF ``a way of making Europe", and the Programa Estatal de Fomento de la Investigaci{\'o}n Cient{\'i}fica y T{\'e}cnica de Excelencia Mar\'{\i}a de Maeztu, grant MDM-2017-0765 and Programa Severo Ochoa del Principado de Asturias (Spain); the Stavros Niarchos Foundation (Greece); the Rachadapisek Sompot Fund for Postdoctoral Fellowship, Chulalongkorn University and the Chulalongkorn Academic into Its 2nd Century Project Advancement Project (Thailand); the Kavli Foundation; the Nvidia Corporation; the SuperMicro Corporation; the Welch Foundation, contract C-1845; and the Weston Havens Foundation (USA).
\end{acknowledgments}

\bibliography{auto_generated}

\providecommand{\href}[2]{#2}\begingroup\raggedright\begin{thebibliography}{10}%
\makeatletter
\providecommand{\hrefCMSnoop }[0]{\@secondoftwo}%
\makeatother
\providecommand{\doi}{\texttt{doi:}\begingroup \urlstyle{tt}\Url}

\bibitem{doi:10.1146/annurev.nucl.55.090704.151502}
\hrefCMSnoop {}{M.~Schmaltz and D.~Tucker-Smith, ``Little {Higgs} theories'',}
  \textit{ Ann. Rev. Nucl. Part. Sci.} \textbf{ 55} (2005) 229,
  \href{http://dx.doi.org/10.1146/annurev.nucl.55.090704.151502}{\doi{10.1146/annurev.nucl.55.090704.151502}},
  \href{http://www.arXiv.org/abs/hep-ph/0502182}{\texttt{arXiv:hep-ph/0502182}}.

\bibitem{PhysRevD.64.035002}
\hrefCMSnoop {}{T.~Appelquist, H.-C. Cheng, and B.~A. Dobrescu, ``Bounds on
  universal extra dimensions'',} \textit{ Phys. Rev. D} \textbf{ 64} (2001)
  035002,
  \href{http://dx.doi.org/10.1103/PhysRevD.64.035002}{\doi{10.1103/PhysRevD.64.035002}},
\href{http://www.arXiv.org/abs/hep-ph/0012100}{\texttt{arXiv:hep-ph/0012100}}.

\bibitem{PhysRevD.11.566}
\hrefCMSnoop {}{R.~N. Mohapatra and J.~C. Pati, ``Left-right gauge symmetry and
  an `isoconjugate' model of {CP} violation'',} \textit{ Phys. Rev. D} \textbf{
  11} (1975) 566,
  \href{http://dx.doi.org/10.1103/PhysRevD.11.566}{\doi{10.1103/PhysRevD.11.566}}.

\bibitem{wpCMS1}
\hrefCMSnoop {}{{CMS Collaboration}, ``{Search for high-mass resonances in
  final states with a lepton and missing transverse momentum at $ \sqrt{s}=13 $
  TeV}'',} \textit{ JHEP} \textbf{ 06} (2018) 128,
  \href{http://dx.doi.org/10.1007/JHEP06(2018)128}{\doi{10.1007/JHEP06(2018)128}},
  \href{http://www.arXiv.org/abs/1803.11133}{\texttt{arXiv:1803.11133}}.

\bibitem{ATLAS:2019lsy}
\hrefCMSnoop {}{{ATLAS Collaboration}, ``{Search for a heavy charged boson in
  events with a charged lepton and missing transverse momentum from pp
  collisions at $\sqrt{s} = 13$ TeV with the ATLAS detector}'',} \textit{ Phys.
  Rev. D} \textbf{ 100} (2019) 052013,
  \href{http://dx.doi.org/10.1103/PhysRevD.100.052013}{\doi{10.1103/PhysRevD.100.052013}},
  \href{http://www.arXiv.org/abs/1906.05609}{\texttt{arXiv:1906.05609}}.

\bibitem{Sirunyan:2018iff}
\hrefCMSnoop {}{{CMS Collaboration}, ``{Search for a heavy resonance decaying
  to a pair of vector bosons in the lepton plus merged jet final state at $
  \sqrt{s}=13 $ TeV}'',} \textit{ JHEP} \textbf{ 05} (2018) 088,
  \href{http://dx.doi.org/10.1007/JHEP05(2018)088}{\doi{10.1007/JHEP05(2018)088}},
\href{http://www.arXiv.org/abs/1802.09407}{\texttt{arXiv:1802.09407}}.

\bibitem{wpAtlas1}
\hrefCMSnoop {}{{ATLAS Collaboration}, ``Search for {WW/WZ} resonance
  production in $\nu$qq final states in pp collisions at {$\sqrt{s}=13 $ TeV}
  with the {ATLAS} detector'',} \textit{ JHEP} \textbf{ 03} (2018) 042,
  \href{http://dx.doi.org/10.1007/JHEP03(2018)042}{\doi{10.1007/JHEP03(2018)042}},
  \href{http://www.arXiv.org/abs/1710.07235}{\texttt{arXiv:1710.07235}}.

\bibitem{wpCMS3}
\hrefCMSnoop {}{{CMS Collaboration}, ``{Searches for $\PWpr$ bosons decaying to
  a top quark and a bottom quark in proton-proton collisions at 13 TeV}'',}
  \textit{ JHEP} \textbf{ 08} (2017) 029,
  \href{http://dx.doi.org/10.1007/JHEP08(2017)029}{\doi{10.1007/JHEP08(2017)029}},
\href{http://www.arXiv.org/abs/1706.04260}{\texttt{arXiv:1706.04260}}.

\bibitem{wpAtlas2}
\hrefCMSnoop {}{{ATLAS Collaboration}, ``{Search for W' $\rightarrow$ tb decays
  in the hadronic final state using pp collisions at $\sqrt{s}=13$ TeV with the
  ATLAS detector}'',} \textit{ Phys. Lett. B} \textbf{ 781} (2018) 327,
  \href{http://dx.doi.org/10.1016/j.physletb.2018.03.036}{\doi{10.1016/j.physletb.2018.03.036}},
\href{http://www.arXiv.org/abs/1801.07893}{\texttt{arXiv:1801.07893}}.

\bibitem{AGASHE2005165}
\hrefCMSnoop {}{K.~Agashe, R.~Contino, and A.~Pomarol, ``{The minimal composite
  Higgs model}'',} \textit{ Nucl. Phys. B} \textbf{ 719} (2005) 165,
  \href{http://dx.doi.org/10.1016/j.nuclphysb.2005.04.035}{\doi{10.1016/j.nuclphysb.2005.04.035}},
\href{http://www.arXiv.org/abs/hep-ph/0412089}{\texttt{arXiv:hep-ph/0412089}}.

\bibitem{Barducci2013}
D.~Barducci\hrefCMSnoop {}{ {et~al.}, ``Exploring {Drell-Yan} signals from the
  {4D} composite {Higgs} model at the {LHC}'',} \textit{ JHEP} \textbf{ 04}
  (2013) 152,
  \href{http://dx.doi.org/10.1007/JHEP04(2013)152}{\doi{10.1007/JHEP04(2013)152}},
  \href{http://www.arXiv.org/abs/1210.2927}{\texttt{arXiv:1210.2927}}.

\bibitem{Barducci2016}
\hrefCMSnoop {}{D.~Barducci and C.~Delaunay, ``{Bounding wide composite vector
  resonances at the LHC}'',} \textit{ JHEP} \textbf{ 02} (2016) 55,
  \href{http://dx.doi.org/10.1007/JHEP02(2016)055}{\doi{10.1007/JHEP02(2016)055}},
  \href{http://www.arXiv.org/abs/1511.01101}{\texttt{arXiv:1511.01101}}.

\bibitem{Vignaroli:2014bpa}
\hrefCMSnoop {}{N.~Vignaroli, ``New {W$'$} signals at the {LHC}'',} \textit{
  Phys. Rev. D} \textbf{ 89} (2014) 095027,
  \href{http://dx.doi.org/10.1103/PhysRevD.89.095027}{\doi{10.1103/PhysRevD.89.095027}},
\href{http://www.arXiv.org/abs/1404.5558}{\texttt{arXiv:1404.5558}}.

\bibitem{vlqs4}
\hrefCMSnoop {}{{CMS Collaboration}, ``{Search for single production of
  vector-like quarks decaying into a b quark and a W boson in proton-proton
  collisions at $\sqrt s =$ 13 TeV}'',} \textit{ Phys. Lett. B} \textbf{ 772}
  (2017) 634,
  \href{http://dx.doi.org/10.1016/j.physletb.2017.07.022}{\doi{10.1016/j.physletb.2017.07.022}},
\href{http://www.arXiv.org/abs/1701.08328}{\texttt{arXiv:1701.08328}}.

\bibitem{vlqs2}
\hrefCMSnoop {}{{CMS Collaboration}, ``{Search for single production of
  vector-like quarks decaying to a b quark and a Higgs boson}'',} \textit{
  JHEP} \textbf{ 06} (2018) 031,
  \href{http://dx.doi.org/10.1007/JHEP06(2018)031}{\doi{10.1007/JHEP06(2018)031}},
\href{http://www.arXiv.org/abs/1802.01486}{\texttt{arXiv:1802.01486}}.

\bibitem{vlqs1}
\hrefCMSnoop {}{{CMS Collaboration}, ``{Search for single production of a
  vector-like T quark decaying to a Z boson and a top quark in proton-proton
  collisions at $\sqrt{s}~=~$13 TeV}'',} \textit{ Phys. Lett. B} \textbf{ 781}
  (2018) 574,
  \href{http://dx.doi.org/10.1016/j.physletb.2018.04.036}{\doi{10.1016/j.physletb.2018.04.036}},
  \href{http://www.arXiv.org/abs/1708.01062}{\texttt{arXiv:1708.01062}}.

\bibitem{vlqs3}
\hrefCMSnoop {}{{ATLAS Collaboration}, ``{Search for pair- and
  single-production of vector-like quarks in final states with at least one $Z$
  boson decaying into a pair of electrons or muons in $pp$ collision data
  collected with the ATLAS detector at $\sqrt{s} = 13$ TeV}'',} \textit{ Phys.
  Rev. D} \textbf{ 98} (2018) 112010,
  \href{http://dx.doi.org/10.1103/PhysRevD.98.112010}{\doi{10.1103/PhysRevD.98.112010}},
  \href{http://www.arXiv.org/abs/1806.10555}{\texttt{arXiv:1806.10555}}.

\bibitem{vlqs5}
\hrefCMSnoop {}{{ATLAS Collaboration}, ``{Search for single production of a
  vector-like $T$ quark decaying into a Higgs boson and top quark with fully
  hadronic final states using the ATLAS detector}'',} \textit{ Phys. Rev. D}
  \textbf{ 105} (1, 2022) 092012,
  \href{http://dx.doi.org/10.1103/PhysRevD.105.092012}{\doi{10.1103/PhysRevD.105.092012}},
  \href{http://www.arXiv.org/abs/2201.07045}{\texttt{arXiv:2201.07045}}.

\bibitem{vlqp1}
\hrefCMSnoop {}{{CMS Collaboration}, ``{Search for pair production of
  vector-like quarks in the bWbW channel from proton-proton collisions at
  $\sqrt{s}~=~$13 TeV}'',} \textit{ Phys. Lett. B} \textbf{ 779} (2018) 82,
  \href{http://dx.doi.org/10.1016/j.physletb.2018.01.077}{\doi{10.1016/j.physletb.2018.01.077}},
  \href{http://www.arXiv.org/abs/1710.01539}{\texttt{arXiv:1710.01539}}.

\bibitem{vlqp2}
\hrefCMSnoop {}{{CMS Collaboration}, ``{Search for vector-like T and B quark
  pairs in final states with leptons at $\sqrt{s} =$ 13 TeV}'',} \textit{ JHEP}
  \textbf{ 08} (2018) 177,
  \href{http://dx.doi.org/10.1007/JHEP08(2018)177}{\doi{10.1007/JHEP08(2018)177}},
  \href{http://www.arXiv.org/abs/1805.04758}{\texttt{arXiv:1805.04758}}.

\bibitem{vlqp4}
\hrefCMSnoop {}{{ATLAS Collaboration}, ``{Combination of the searches for
  pair-produced vector-like partners of the third-generation quarks at
  $\sqrt{s} =$ 13 TeV with the ATLAS detector}'',} \textit{ Phys. Rev. Lett.}
  \textbf{ 121} (2018) 211801,
  \href{http://dx.doi.org/10.1103/PhysRevLett.121.211801}{\doi{10.1103/PhysRevLett.121.211801}},
  \href{http://www.arXiv.org/abs/1808.02343}{\texttt{arXiv:1808.02343}}.

\bibitem{vlqp3}
\hrefCMSnoop {}{{CMS Collaboration}, ``{A search for bottom-type, vector-like
  quark pair production in a fully hadronic final state in proton-proton
  collisions at $\sqrt{s} =$ 13 TeV}'',} \textit{ Phys. Rev. D} \textbf{ 102}
  (2020) 112004,
  \href{http://dx.doi.org/10.1103/PhysRevD.102.112004}{\doi{10.1103/PhysRevD.102.112004}},
  \href{http://www.arXiv.org/abs/2008.09835}{\texttt{arXiv:2008.09835}}.

\bibitem{CMS:2021xjt}
\hrefCMSnoop {}{{CMS Collaboration}, ``{Precision luminosity measurement in
  proton-proton collisions at $\sqrt{s} =$ 13 TeV in 2015 and 2016 at CMS}'',}
  \textit{ Eur. Phys. J. C} \textbf{ 81} (2021) 800,
  \href{http://dx.doi.org/10.1140/epjc/s10052-021-09538-2}{\doi{10.1140/epjc/s10052-021-09538-2}},
  \href{http://www.arXiv.org/abs/2104.01927}{\texttt{arXiv:2104.01927}}.

\bibitem{CMS-PAS-LUM-17-004}
\href {https://cds.cern.ch/record/2621960/}{{{CMS}} Collaboration, ``{CMS}
  luminosity measurement for the 2017 data-taking period at $\sqrt{s}$ = 13
  {TeV}'',} CMS Physics Analysis Summary CMS-PAS-LUM-17-004, 2018.

\bibitem{CMS-PAS-LUM-18-002}
\href {https://cds.cern.ch/record/2676164/}{{{CMS}} Collaboration, ``{CMS}
  luminosity measurement for the 2018 data-taking period at $\sqrt{s}$ = 13
  {TeV}'',} CMS Physics Analysis Summary CMS-PAS-LUM-18-002, 2019.

\bibitem{hepdata}
\hrefCMSnoop {}{}{HEPD}ata record for this analysis, 2022.
\newblock
  \href{http://dx.doi.org/10.17182/hepdata.127138}{\doi{10.17182/hepdata.127138}}.

\bibitem{Sirunyan:2020zal}
\hrefCMSnoop {}{{CMS Collaboration}, ``{Performance of the CMS Level-1 trigger
  in proton-proton collisions at $\sqrt{s} = 13$\,TeV}'',} \textit{ JINST}
  \textbf{ 15} (2020) P10017,
  \href{http://dx.doi.org/10.1088/1748-0221/15/10/P10017}{\doi{10.1088/1748-0221/15/10/P10017}},
  \href{http://www.arXiv.org/abs/2006.10165}{\texttt{arXiv:2006.10165}}.

\bibitem{Khachatryan:2016bia}
\hrefCMSnoop {}{{CMS Collaboration}, ``{The CMS trigger system}'',} \textit{
  JINST} \textbf{ 12} (2017) P01020,
  \href{http://dx.doi.org/10.1088/1748-0221/12/01/P01020}{\doi{10.1088/1748-0221/12/01/P01020}},
\href{http://www.arXiv.org/abs/1609.02366}{\texttt{arXiv:1609.02366}}.

\bibitem{Chatrchyan:2008zzk}
\hrefCMSnoop {}{{CMS Collaboration}, ``The {CMS} experiment at the {CERN}
  {LHC}'',} \textit{ JINST} \textbf{ 3} (2008) S08004,
  \href{http://dx.doi.org/10.1088/1748-0221/3/08/S08004}{\doi{10.1088/1748-0221/3/08/S08004}}.

\bibitem{CMS-PRF-14-001}
\hrefCMSnoop {}{{CMS Collaboration}, ``Particle-flow reconstruction and global
  event description with the {CMS} detector'',} \textit{ JINST} \textbf{ 12}
  (2017) P10003,
  \href{http://dx.doi.org/10.1088/1748-0221/12/10/P10003}{\doi{10.1088/1748-0221/12/10/P10003}},
\href{http://www.arXiv.org/abs/1706.04965}{\texttt{arXiv:1706.04965}}.

\bibitem{Cacciari:2008gp}
\hrefCMSnoop {}{M.~Cacciari, G.~P. Salam, and G.~Soyez, ``The anti-$\kt$ jet
  clustering algorithm'',} \textit{ JHEP} \textbf{ 04} (2008) 063,
  \href{http://dx.doi.org/10.1088/1126-6708/2008/04/063}{\doi{10.1088/1126-6708/2008/04/063}},
  \href{http://www.arXiv.org/abs/0802.1189}{\texttt{arXiv:0802.1189}}.

\bibitem{Cacciari:2011ma}
\hrefCMSnoop {}{M.~Cacciari, G.~P. Salam, and G.~Soyez, ``{FastJet} user
  manual'',} \textit{ Eur. Phys. J. C} \textbf{ 72} (2012) 1896,
  \href{http://dx.doi.org/10.1140/epjc/s10052-012-1896-2}{\doi{10.1140/epjc/s10052-012-1896-2}},
\href{http://www.arXiv.org/abs/1111.6097}{\texttt{arXiv:1111.6097}}.

\bibitem{Bertolini2014}
\hrefCMSnoop {}{D.~Bertolini, P.~Harris, M.~Low, and N.~Tran, ``{Pileup per
  particle identification}'',} \textit{ JHEP} \textbf{ 10} (2014) 059,
  \href{http://dx.doi.org/10.1007/JHEP10(2014)059}{\doi{10.1007/JHEP10(2014)059}},
\href{http://www.arXiv.org/abs/1407.6013}{\texttt{arXiv:1407.6013}}.

\bibitem{Khachatryan:2016kdb}
\hrefCMSnoop {}{{CMS Collaboration}, ``Jet energy scale and resolution in the
  {CMS} experiment in pp collisions at 8 {TeV}'',} \textit{ JINST} \textbf{ 12}
  (2017) P02014,
  \href{http://dx.doi.org/10.1088/1748-0221/12/02/P02014}{\doi{10.1088/1748-0221/12/02/P02014}},
\href{http://www.arXiv.org/abs/1607.03663}{\texttt{arXiv:1607.03663}}.

\bibitem{Nason:2004rx}
\hrefCMSnoop {}{P.~Nason, ``A new method for combining {NLO QCD} with shower
  {M}onte {C}arlo algorithms'',} \textit{ JHEP} \textbf{ 11} (2004) 040,
  \href{http://dx.doi.org/10.1088/1126-6708/2004/11/040}{\doi{10.1088/1126-6708/2004/11/040}},
\href{http://www.arXiv.org/abs/hep-ph/0409146}{\texttt{arXiv:hep-ph/0409146}}.

\bibitem{Powheg}
\hrefCMSnoop {}{S.~Frixione, P.~Nason, and C.~Oleari, ``Matching {NLO QCD}
  computations with parton shower simulations: the {POWHEG} method'',} \textit{
  JHEP} \textbf{ 11} (2007) 070,
  \href{http://dx.doi.org/10.1088/1126-6708/2007/11/070}{\doi{10.1088/1126-6708/2007/11/070}},
\href{http://www.arXiv.org/abs/0709.2092}{\texttt{arXiv:0709.2092}}.

\bibitem{Alioli:2010xd}
\hrefCMSnoop {}{S.~Alioli, P.~Nason, C.~Oleari, and E.~Re, ``A general
  framework for implementing {NLO} calculations in shower {M}onte {C}arlo
  programs: the {POWHEG BOX}'',} \textit{ JHEP} \textbf{ 06} (2010) 043,
  \href{http://dx.doi.org/10.1007/JHEP06(2010)043}{\doi{10.1007/JHEP06(2010)043}},
\href{http://www.arXiv.org/abs/1002.2581}{\texttt{arXiv:1002.2581}}.

\bibitem{Alioli:2009je}
\hrefCMSnoop {}{S.~Alioli, P.~Nason, C.~Oleari, and E.~Re, ``{NLO single-top
  production matched with shower in POWHEG: $s$- and $t$-channel
  contributions}'',} \textit{ JHEP} \textbf{ 09} (2009) 111,
  \href{http://dx.doi.org/10.1088/1126-6708/2009/09/111}{\doi{10.1088/1126-6708/2009/09/111}},
  \href{http://www.arXiv.org/abs/0907.4076}{\texttt{arXiv:0907.4076}}.
[Erratum: \DOI{10.1007/JHEP02(2010)011}].

\bibitem{Frederix:2012dh}
\hrefCMSnoop {}{R.~Frederix, E.~Re, and P.~Torrielli, ``{Single-top}
  $t$-channel hadroproduction in the four-flavour scheme with {POWHEG} and
  {aMC@NLO}'',} \textit{ JHEP} \textbf{ 09} (2012) 130,
  \href{http://dx.doi.org/10.1007/JHEP09(2012)130}{\doi{10.1007/JHEP09(2012)130}},
\href{http://www.arXiv.org/abs/1207.5391}{\texttt{arXiv:1207.5391}}.

\bibitem{Alioli:2011as}
\hrefCMSnoop {}{S.~Alioli, S.-O. Moch, and P.~Uwer, ``{Hadronic top-quark
  pair-production with one jet and parton showering}'',} \textit{ JHEP}
  \textbf{ 01} (2012) 137,
  \href{http://dx.doi.org/10.1007/JHEP01(2012)137}{\doi{10.1007/JHEP01(2012)137}},
  \href{http://www.arXiv.org/abs/1110.5251}{\texttt{arXiv:1110.5251}}.

\bibitem{Alwall:2014hca}
J.~Alwall\hrefCMSnoop {}{ {et~al.}, ``{The automated computation of tree-level
  and next-to-leading order differential cross sections, and their matching to
  parton shower simulations}'',} \textit{ JHEP} \textbf{ 07} (2014) 079,
  \href{http://dx.doi.org/10.1007/JHEP07(2014)079}{\doi{10.1007/JHEP07(2014)079}},
\href{http://www.arXiv.org/abs/1405.0301}{\texttt{arXiv:1405.0301}}.

\bibitem{Alwall:2007fs}
J.~Alwall\hrefCMSnoop {}{ {et~al.}, ``Comparative study of various algorithms
  for the merging of parton showers and matrix elements in hadronic
  collisions'',} \textit{ Eur. Phys. J. C} \textbf{ 53} (2008) 473,
  \href{http://dx.doi.org/10.1140/epjc/s10052-007-0490-5}{\doi{10.1140/epjc/s10052-007-0490-5}},
\href{http://www.arXiv.org/abs/0706.2569}{\texttt{arXiv:0706.2569}}.

\bibitem{Ball:2014uwa}
\hrefCMSnoop {}{{NNPDF} Collaboration, ``{Parton distributions for the LHC Run
  II}'',} \textit{ JHEP} \textbf{ 04} (2015) 040,
  \href{http://dx.doi.org/10.1007/JHEP04(2015)040}{\doi{10.1007/JHEP04(2015)040}},
  \href{http://www.arXiv.org/abs/1410.8849}{\texttt{arXiv:1410.8849}}.

\bibitem{Ball:2017nwa}
\hrefCMSnoop {}{{NNPDF} Collaboration, ``Parton distributions from
  high-precision collider data'',} \textit{ Eur. Phys. J. C} \textbf{ 77}
  (2017) 663,
  \href{http://dx.doi.org/10.1140/epjc/s10052-017-5199-5}{\doi{10.1140/epjc/s10052-017-5199-5}},
\href{http://www.arXiv.org/abs/1706.00428}{\texttt{arXiv:1706.00428}}.

\bibitem{Sjostrand:2014zea}
T.~Sj{\"o}strand\hrefCMSnoop {}{ {et~al.}, ``An introduction to {PYTHIA}
  8.2'',} \textit{ Comput. Phys. Commun.} \textbf{ 191} (2015) 159,
  \href{http://dx.doi.org/10.1016/j.cpc.2015.01.024}{\doi{10.1016/j.cpc.2015.01.024}},
\href{http://www.arXiv.org/abs/1410.3012}{\texttt{arXiv:1410.3012}}.

\bibitem{CMS-PAS-TOP-16-021}
\href {https://cds.cern.ch/record/2235192}{{CMS Collaboration},
  ``Investigations of the impact of the parton shower tuning in {Pythia} 8 in
  the modelling of $\mathrm{t\overline{t}}$ at $\sqrt{s}=8$ and 13 {TeV}'',}
  CMS Physics Analysis Summary CMS-PAS-TOP-16-021, 2016.

\bibitem{Khachatryan:2015pea}
\hrefCMSnoop {}{{CMS Collaboration}, ``Event generator tunes obtained from
  underlying event and multiparton scattering measurements'',} \textit{ Eur.
  Phys. J. C} \textbf{ 76} (2016) 155,
  \href{http://dx.doi.org/10.1140/epjc/s10052-016-3988-x}{\doi{10.1140/epjc/s10052-016-3988-x}},
\href{http://www.arXiv.org/abs/1512.00815}{\texttt{arXiv:1512.00815}}.

\bibitem{CMS:2019csb}
\hrefCMSnoop {}{{CMS Collaboration}, ``Extraction and validation of a new set
  of {CMS} {PYTHIA8} tunes from underlying-event measurements'',} \textit{ Eur.
  Phys. J. C} \textbf{ 80} (2020) 4,
  \href{http://dx.doi.org/10.1140/epjc/s10052-019-7499-4}{\doi{10.1140/epjc/s10052-019-7499-4}},
  \href{http://www.arXiv.org/abs/1903.12179}{\texttt{arXiv:1903.12179}}.

\bibitem{g4c}
\hrefCMSnoop {}{{GEANT4} Collaboration, ``{\GEANTfour}---a simulation
  toolkit'',} \textit{ Nucl. Instrum. Meth.} \textbf{ 506} (2003) 250,
  \href{http://dx.doi.org/10.1016/S0168-9002(03)01368-8}{\doi{10.1016/S0168-9002(03)01368-8}}.

\bibitem{Sirunyan:2020lcu}
\hrefCMSnoop {}{{CMS Collaboration}, ``{Identification of heavy, energetic,
  hadronically decaying particles using machine-learning techniques}'',}
  \textit{ JINST} \textbf{ 15} (2020) P06005,
  \href{http://dx.doi.org/10.1088/1748-0221/15/06/P06005}{\doi{10.1088/1748-0221/15/06/P06005}},
  \href{http://www.arXiv.org/abs/2004.08262}{\texttt{arXiv:2004.08262}}.

\bibitem{Dasgupta2013}
\hrefCMSnoop {}{M.~Dasgupta, A.~Fregoso, S.~Marzani, and G.~P. Salam,
  ``{Towards an understanding of jet substructure}'',} \textit{ JHEP} \textbf{
  09} (2013) 029,
  \href{http://dx.doi.org/10.1007/JHEP09(2013)029}{\doi{10.1007/JHEP09(2013)029}},
\href{http://www.arXiv.org/abs/1307.0007}{\texttt{arXiv:1307.0007}}.

\bibitem{Larkoski:2014wba}
\hrefCMSnoop {}{A.~J. Larkoski, S.~Marzani, G.~Soyez, and J.~Thaler, ``Soft
  drop'',} \textit{ JHEP} \textbf{ 05} (2014) 146,
  \href{http://dx.doi.org/10.1007/JHEP05(2014)146}{\doi{10.1007/JHEP05(2014)146}},
  \href{http://www.arXiv.org/abs/1402.2657}{\texttt{arXiv:1402.2657}}.

\bibitem{PhysRevD.97.072006}
\hrefCMSnoop {}{{CMS Collaboration}, ``{Search for massive resonances decaying
  into WW, WZ, ZZ, qW, and qZ with dijet final states at $\sqrt{s}=13\text{
  }\text{ }\mathrm{TeV}$}'',} \textit{ Phys. Rev. D} \textbf{ 97} (2018)
  072006,
  \href{http://dx.doi.org/10.1103/PhysRevD.97.072006}{\doi{10.1103/PhysRevD.97.072006}},
\href{http://www.arXiv.org/abs/1708.05379}{\texttt{arXiv:1708.05379}}.

\bibitem{CMS-PAS-JME-16-003}
\href {http://cds.cern.ch/record/2256875}{{CMS Collaboration}, ``Jet algorithms
  performance in 13 {TeV} data'',} CMS Physics Analysis Summary
  CMS-PAS-JME-16-003, 2017.

\bibitem{Sirunyan:2017ezt}
\hrefCMSnoop {}{{CMS Collaboration}, ``{Identification of heavy-flavour jets
  with the CMS detector in pp collisions at 13 TeV}'',} \textit{ JINST}
  \textbf{ 13} (2018) P05011,
  \href{http://dx.doi.org/10.1088/1748-0221/13/05/P05011}{\doi{10.1088/1748-0221/13/05/P05011}},
\href{http://www.arXiv.org/abs/1712.07158}{\texttt{arXiv:1712.07158}}.

\bibitem{Thaler:2011gf}
\hrefCMSnoop {}{J.~Thaler and K.~Van~Tilburg, ``Maximizing boosted top
  identification by minimizing {$N$}-subjettiness'',} \textit{ JHEP} \textbf{
  02} (2012) 093,
  \href{http://dx.doi.org/10.1007/JHEP02(2012)093}{\doi{10.1007/JHEP02(2012)093}},
\href{http://www.arXiv.org/abs/1108.2701}{\texttt{arXiv:1108.2701}}.

\bibitem{Bols:2020bkb}
E.~Bols\hrefCMSnoop {}{ {et~al.}, ``{Jet flavour classification using
  DeepJet}'',} \textit{ JINST} \textbf{ 15} (2020) P12012,
  \href{http://dx.doi.org/10.1088/1748-0221/15/12/P12012}{\doi{10.1088/1748-0221/15/12/P12012}},
  \href{http://www.arXiv.org/abs/2008.10519}{\texttt{arXiv:2008.10519}}.

\bibitem{CMS-DP-2018-058}
\href {http://cds.cern.ch/record/2646773}{{CMS Collaboration}, ``{Performance
  of the DeepJet b tagging algorithm using 41.9/fb of data from proton-proton
  collisions at 13 TeV with Phase 1 CMS detector}'',} CMS Detector Performance
  Note CMS-DP-2018-058, 2018.

\bibitem{PhysRevD.95.092001}
\hrefCMSnoop {}{{CMS Collaboration}, ``{Measurement of differential cross
  sections for top quark pair production using the lepton+jets final state in
  proton-proton collisions at 13 TeV}'',} \textit{ Phys. Rev. D} \textbf{ 95}
  (2017) 092001,
  \href{http://dx.doi.org/10.1103/PhysRevD.95.092001}{\doi{10.1103/PhysRevD.95.092001}},
\href{http://www.arXiv.org/abs/1610.04191}{\texttt{arXiv:1610.04191}}.

\bibitem{Conway:2011in}
\hrefCMSnoop {}{J.~S. Conway, ``{Incorporating nuisance parameters in
  likelihoods for multisource spectra}'',} in \textit{ {Proceedings, PHYSTAT
  2011 workshop on statistical issues related to discovery claims in search
  experiments and unfolding, CERN,Geneva, Switzerland 17-20 January 2011}},
  p.~115.
\newblock 2011.
\newblock \href{http://www.arXiv.org/abs/1103.0354}{\texttt{arXiv:1103.0354}}.
\newblock
\href{http://dx.doi.org/10.5170/CERN-2011-006.115}{\doi{10.5170/CERN-2011-006.115}}.

\bibitem{Sirunyan:2018nqx}
\hrefCMSnoop {}{{CMS Collaboration}, ``{Measurement of the inelastic
  proton-proton cross section at $ \sqrt{s}=13 $ TeV}'',} \textit{ JHEP}
  \textbf{ 07} (2018) 161,
  \href{http://dx.doi.org/10.1007/JHEP07(2018)161}{\doi{10.1007/JHEP07(2018)161}},
\href{http://www.arXiv.org/abs/1802.02613}{\texttt{arXiv:1802.02613}}.

\bibitem{Botje:2011sn}
\hrefCMSnoop {}{M.~Botje {et~al.}, ``{The PDF4LHC working group interim
  recommendations}'',} 2011.
\href{http://www.arXiv.org/abs/1101.0538}{\texttt{arXiv:1101.0538}}.

\bibitem{Alekhin:2011sk}
\hrefCMSnoop {}{S.~Alekhin {et~al.}, ``{The PDF4LHC working group interim
  report}'',} 2011.
\href{http://www.arXiv.org/abs/1101.0536}{\texttt{arXiv:1101.0536}}.

\bibitem{BARLOW1993219}
\hrefCMSnoop {}{R.~J. Barlow and C.~Beeston, ``{Fitting using finite Monte
  Carlo samples}'',} \textit{ Comput. Phys. Commun.} \textbf{ 77} (1993) 219,
  \href{http://dx.doi.org/10.1016/0010-4655(93)90005-W}{\doi{10.1016/0010-4655(93)90005-W}}.

\bibitem{Junk:1999kv}
\hrefCMSnoop {}{T.~Junk, ``{Confidence level computation for combining searches
  with small statistics}'',} \textit{ Nucl. Instrum. Meth. A} \textbf{ 434}
  (1999) 435,
  \href{http://dx.doi.org/10.1016/S0168-9002(99)00498-2}{\doi{10.1016/S0168-9002(99)00498-2}},
  \href{http://www.arXiv.org/abs/hep-ex/9902006}{\texttt{arXiv:hep-ex/9902006}}.

\bibitem{Read:2002hq}
\hrefCMSnoop {}{A.~L. Read, ``{Presentation of search results: The \CLs
  technique}'',} \textit{ J. Phys. G} \textbf{ 28} (2002) 2693,
  \href{http://dx.doi.org/10.1088/0954-3899/28/10/313}{\doi{10.1088/0954-3899/28/10/313}}.

\bibitem{Cowan:2010js}
\hrefCMSnoop {}{G.~Cowan, K.~Cranmer, E.~Gross, and O.~Vitells, ``{Asymptotic
  formulae for likelihood-based tests of new physics}'',} \textit{ Eur. Phys.
  J. C} \textbf{ 71} (2011) 1554,
  \href{http://dx.doi.org/10.1140/epjc/s10052-011-1554-0}{\doi{10.1140/epjc/s10052-011-1554-0}},
  \href{http://www.arXiv.org/abs/1007.1727}{\texttt{arXiv:1007.1727}}.
  [Erratum: \DOI{10.1140/epjc/s10052-013-2501-z}].

\end{thebibliography}\endgroup
\cleardoublepage \appendix\section{The CMS Collaboration \label{app:collab}}\begin{sloppypar}\hyphenpenalty=5000\widowpenalty=500\clubpenalty=5000\cmsinstitute{Yerevan~Physics~Institute, Yerevan, Armenia}
A.~Tumasyan
\cmsinstitute{Institut~f\"{u}r~Hochenergiephysik, Vienna, Austria}
W.~Adam\cmsorcid{0000-0001-9099-4341}, J.W.~Andrejkovic, T.~Bergauer\cmsorcid{0000-0002-5786-0293}, S.~Chatterjee\cmsorcid{0000-0003-2660-0349}, M.~Dragicevic\cmsorcid{0000-0003-1967-6783}, A.~Escalante~Del~Valle\cmsorcid{0000-0002-9702-6359}, R.~Fr\"{u}hwirth\cmsAuthorMark{1}, M.~Jeitler\cmsAuthorMark{1}\cmsorcid{0000-0002-5141-9560}, N.~Krammer, L.~Lechner\cmsorcid{0000-0002-3065-1141}, D.~Liko, I.~Mikulec, P.~Paulitsch, F.M.~Pitters, J.~Schieck\cmsAuthorMark{1}\cmsorcid{0000-0002-1058-8093}, R.~Sch\"{o}fbeck\cmsorcid{0000-0002-2332-8784}, D.~Schwarz, S.~Templ\cmsorcid{0000-0003-3137-5692}, W.~Waltenberger\cmsorcid{0000-0002-6215-7228}, C.-E.~Wulz\cmsAuthorMark{1}\cmsorcid{0000-0001-9226-5812}
% ...
\vskip\cmsinstskip
% ...
1:~Also at TU Wien, Wien, Austria\\
% ...
\end{sloppypar}
\end{document}